\documentclass[aps,pre,preprint,amsmath,amssymb,groupedaddress,showpacs,longbibliography]{revtex4-1}

\usepackage{graphicx}% Include figure files
\newcommand{\pd}[2]{\frac{\partial #1}{\partial #2}}

\usepackage{color}
\definecolor{gray}{rgb}{0.5,0.5,0.5}

\begin{document}
%\preprint{}

\title{Passage of a shock wave through inhomogeneous media and its impact 
on a gas bubble deformation}


\author{A. F. Nowakowski}
\email[E-mail:]{A.F.Nowakowski@sheffield.ac.uk}

\author{A. Ballil}
\altaffiliation[Present address: ]{Department of Mechanical Engineering, 
 University of Benghazi, Benghazi, Libya}

\author{F. C. G. A. Nicolleau}
\affiliation{Sheffield Fluid Mechanics Group SFMG, 
Department of Mechanical Engineering, 
University of Sheffield, Mappin Building, 
Mappin Street, Sheffield, S1 3JD, United Kingdom}


\date{\today}

\begin{abstract}
The paper investigates shock-induced vortical flows within inhomogeneous 
media of nonuniform thermodynamic properties. Numerical simulations are 
performed using an Eularian type mathematical model for compressible 
multi-component flow problems. The model, which accounts for pressure 
non-equilibrium and applies different equations of state for individual 
flow components, shows excellent capabilities for the resolution of interfaces 
separating compressible fluids as well as for capturing the baroclinic source 
of vorticity generation. The developed finite volume Godunov type 
computational approach is equipped with an approximate Riemann solver for 
calculating fluxes and handles numerically diffused zones at flow component 
interfaces. The computations are performed for various initial conditions 
and are compared with available experimental data. The initial conditions 
promoting a shock-bubble interaction process include: weak to high 
planar shock waves with a Mach number ranging from $1.2$ to $3$ and isolated 
cylindrical bubble inhomogeneities of helium, argon, nitrogen, krypton and 
sulphur hexafluoride. The numerical results reveal the characteristic 
features of the evolving flow topology. The impulsively generated flow 
perturbations are dominated by the reflection and refraction of the shock, 
the compression and acceleration as well as the vorticity generation within 
the medium. The study is further extended to investigate the influence
of the ratio of the heat capacities on the interface deformation.
\end{abstract}
\pacs{47.40.Nm, 47.55.-t, 47.11.-j, 47.32.-y}
%\keywords{}

\maketitle

\section{\label{s_one}Introduction}
%%%%%%%%%%%%%%%%%%%%%%%%%%%%%%%%%%%%%
Compressible multi-component flows with low to high density ratios between
components are involved in various physical phenomena and many industrial
applications. Some important examples are inertial confinement fusion 
(ICF)~\cite{Aglitskiy2010}, rapid and efficient mixing of fuel and oxidizer
in supersonic combustion, primary fuel atomization in aircraft engines and 
droplet breakup. A proper understanding of these flows requires studying 
the evolution and creation of interfaces resulting from the interaction 
of a shock wave with the environment of inhomogeneous gases. The diverse 
flow patterns and the dynamical interaction of gas phases at the interface 
could cause several physical processes to occur simultaneously. These include 
shock acceleration or refraction, vorticity generation and its transport, 
and consequently shock-induced turbulence. The mechanism of these processes 
is related to the strength and pattern of the propagating shock waves during 
the short time of their encounter with the surface's curvatures between flow 
components and inherently to the difference in the acoustic impedance at the 
components' interfaces. The recent review paper~\cite{Ranjan2011} 
provides an excellent description of various possible phenomena occurring 
during the shock bubble interaction process.  When, as a result of the passage 
of a shock wave, an interface between fluid components is impulsively 
accelerated, the development of a so called Richtmyer-Meshkov instability
(RMI)~\cite{Brouillette2002} can be observed. The instability, which directly results
from the amplification of perturbations at the interface, is due to baroclinic 
vorticity generation as a consequence of the misalignment of the pressure 
gradient of the shock and the local density gradient across the interface. 
This is a complex phenomenon constituting a challenging task to investigate 
either experimentally or numerically as the derivation of a mathematical model
for this problem is not straightforward. 

The nature of the impulsively generated perturbations at the interface of 
two-component compressible flows has been studied experimentally using 
idealised configurations. The interaction of a planar shock wave with 
a cylinder or a sphere is a typical physical arrangement that has received 
attention. However, the measurement of the entire velocity, 
density and pressure fields for a large selection of physical scales and 
interface geometries remains an enormous experimental challenge. 
	
The first key work to monitor the interaction between a plane shock wave 
and a single gas bubble was presented in~\cite{Haas1987}. The shadowgraph 
photography technique was utilised to visualise a wave front geometry and the 
deformation of the gas bubble volume. The distortion of a spherical bubble 
impacted by a plane shock wave was later examined 
in~\cite{Layes2003, Layes2007, Layes2009} using a high speed 
rotating camera shadowgraph system and in~\cite{Zhai2011, Si2012} by means 
of the high speed schlieren photography with higher time resolutions. All 
these experiments were conducted in horizontal shock tubes characterised 
by a Mach number smaller than $1.7$. 
Other laser based shock-tube experiments 
in~\cite{Ranjan2007, Ranjan2008, Ranjan2011, Haehn2012} 
covered a wider selection of Mach numbers and provided qualitative and 
quantitative data for the shock-bubble interaction within the Mach number 
range of $1.1-3.5$.  A similar geometry was investigated in~\cite{Tomkins2008} 
to find a mixing mechanism in a shock-induced instability flow. Although at 
present it is possible to consider experiments with a higher Mach number 
by building laboratory facilities based on modern laser technologies, 
such tests still remain rather difficult and expensive to conduct.
	
Therefore the development of numerical techniques for these types of 
applications seems to be an ideal alternative to provide reasonable 
results at a significantly lower cost. A shock-capturing upwind finite 
difference numerical method has been utilised to solve the compressible 
Euler equations for two species in an axisymmetric two-dimensional case 
of planar shock interacting with a bubble~\cite{Zabusky1998}. The evolution
of upstream and downstream complex wave patterns and the appearance of vortex
rings were resolved in this study. The experiment in~\cite{Haas1987}, in 
which a shock wave with a Mach number $1.22$ hits a helium bubble, 
has inspired several other authors~\cite{Picone1988, Quirk1996, Bagabir2001, 
Banks2007, Chang2007, Terashima2009, Hejazialhosseini2010,  Shukla2010, 
So2012, Franquet2012} who adopted this experiment to demonstrate the 
performance of the numerical techniques they developed. In the majority of the 
cases the authors used the Euler equations to simulate the experiment and 
the interface reconstruction was the major task. 
For example~\cite{Terashima2009} proposed to use the front tracking/ghost 
fluid method to capture  fluid interface minimizing at the same time the 
smearing of discontinuous variables. 
In another development~\cite{Niederhaus2008} the two-dimensional simulations 
of the shock-bubble interaction were extended to three spatial dimensions 
and high Mach numbers using the volume-of-fluid (VOF) method as the numerical
approach. The authors considered fourteen different scenarios, including four 
gas pairings by using a numerical algorithm solving the same system of partial
differential equations for each of the two constituent species with an 
additional numerical scheme for the local interface reconstruction. 
The 2D VOF method was also used in~\cite{So2012}. A viscous approach, 
but without accounting for turbulence, was adopted in the numerical 
study~\cite{Giordano2006} which reproduced different experiments performed
in~\cite{Layes2003}. 
	
The majority of numerical simulations are based on the mixture Euler equations
supplemented by two species conservation equations in order to build a reasonable 
equation of state parameters at the interface 
(see example~\cite{Quirk1996, Bagabir2001}). By contrast, the Arbitrary 
Lagrangian Eulerian methods or Front Tracking Methods~\cite{Terashima2009} 
consider multi-material interfaces as genuine sharp non-smeared 
discontinuities. These methods are less flexible when dealing with situations 
of large interface deformations and topological changes. 
	
The mathematical model advocated by the authors of this paper, is based on 
a different point of view, and while considering the equations for immiscible 
fluids, does not require the explicit application of boundary conditions at 
the interface. The system of equations can be derived from 
the Baer and Nunziato model~\cite{Baer1986}. However, in contrast 
to~\cite{Franquet2012}, who used the original~\cite{Baer1986} formulation 
to investigate shock-bubble interaction, the equations presented in this paper
are considered in an asymptotic limit of the velocity relaxation time of the
model~\cite{Baer1986}. This so called 
\textquotedblleft six-equation model\textquotedblright 
was derived for the first time in~\cite{Kapila2001} and was further 
investigated in~\cite{Saurel2009}. 
In the latter reference~\cite{Saurel2009} the authors showed that the 
non-monotonic behaviour of the sound speed which causes errors in the 
transmission of waves across interfaces can be circumvented  by restoring 
the effects of pressure non-equilibrium in the equation of the volume fraction 
evolution by using two pressures and their associated pressure relaxation 
terms.  The six-equation model takes advantage of the inherent numerical 
diffusion at the interface as the necessary condition for interface capturing 
and avoids the spurious pressure oscillations that frequently occur at the 
multi-fluid interfaces. Furthermore, and what is of key importance here, 
the model can naturally handle complex topological changes. The other 
attractive and desired features of this model could be summarised as
follows:  the ability to simulate the dynamical creation and the evolution
of interfaces, the numerical implementation with a single solver for a system
of unified conservation equations and the ability to use different equations 
of state and hence different heat capacity ratios for individual flow components. 
	
This paper investigates the two-dimensional flow of a shock wave encountering
circular inhomogeneities. It presents a numerical study of the interaction 
of weak to high Mach number waves with an inhomogeneous medium containing 
a gas bubble. 
The inherent features of such flow composition are density jumps across the 
interface. The study concentrates on the early phases of the interaction 
process. The purpose is to consider the influence of both the Atwood number 
and the shock wave Mach number on the deformation of the gas bubbles and 
the associated production of a vorticity field. The physical behaviour of the 
gas bubbles is monitored using a newly developed numerical algorithm which 
has been built to solve the six equation model. The considered Atwood numbers 
are within the range ($-0.8 \leqslant A \leqslant 0.7$) and the shock celerity 
covers the range ($1.2\leqslant \textit{Ma} \leqslant 3$). 
	
The outline of the paper is as follows: section~\ref{sec:2} gives a brief 
introduction to the two-component flow governing equations. 
Then in section~\ref{sec:3} the numerical procedures to solve the system 
are described. The main focus of this paper is section~\ref{sec:4}, which 
presents the results of the computational work. 
First, two independent experimental investigations described 
in~\cite{Layes2007} and~\cite{Zhai2011} are used for the interface evolution 
validation. In the case of~\cite{Layes2007}, a shock wave ($\textit{Ma} = 1.5$) 
interacts with three different air/gas configurations which are air/helium (He), 
air/nitrogen (N$_ {2}$) and air/krypton (Kr). In the case of ~\cite{Zhai2011} 
a shock wave ($\textit{Ma} = 1.5$) interacts with a sulphur hexafluoride (SF$_6$)
bubble.
Second, the study is extended to account for the different 
gas pairings in an attempt to evaluate the effect of the Atwood number 
on the complex pattern of the gas bubbles evolution. Third, the effect of 
the Mach number on the interface evolution is investigated for all cases with 
the intention to discuss and quantify the production of vorticity resulting 
from the passage of the shock. Finally, the investigation of the influence
of the ratio of the heat capacities on the interface deformation is made. 
The conclusions are drawn in section~\ref{sec:5}.

\section{\label{sec:2}Mathematical model}
A two-component compressible flow model is considered. The model consists 
of two separate, identifiable and interpenetrating continua that are in 
thermodynamic non-equilibrium with each other. In its one-dimensional 
mathematical framework the model, first derived in~\cite{Kapila2001}, 
consists of six partial differential equations. It constitutes a reduced form
of the more general seven equation model~\cite{Baer1986}. The one dimensional 
equations of the model are: a statistical volume fraction equation, two 
continuity equations, one momentum equation and two energy equations. 
It differs from more popular models which rely on instantaneous pressure 
equilibrium between the two flow components or phases~\cite{Murrone2005}. 
The original six-equation model can be expressed in one dimensional space, 
as follows: 
\begin{eqnarray}\label{eq:1}
&&\pd{\alpha_1}{t}+u\pd{\alpha_1}{x} = \mu(p_1-p_2),\nonumber\\
&&\pd{\alpha_1\rho_1}{t}+\pd{\alpha_1\rho_1u}{x} = 0,\nonumber\\
&&\pd{\alpha_2\rho_2}{t}+\pd{\alpha_2\rho_2u}{x} = 0,\nonumber\\
&&\pd{\rho u}{t}+\pd{[\rho u^2+(\alpha_1p_1+\alpha_2p_2)]}{x} = 0,\\
&&\pd{\alpha_1\rho_1 e_1}{t}+\pd{\alpha_1\rho_1 e_1u}{x}+\alpha_1p_1\pd{u}{x} = P_i \mu(p_1-p_2),\nonumber\\
&&\pd{\alpha_2\rho_2 e_2}{t}+\pd{\alpha_2\rho_2 e_2u}{x}+\alpha_2p_2\pd{u}{x} = -P_i \mu(p_1-p_2),\nonumber
\end{eqnarray}
where $\alpha_k$, $\rho_k$, $p_k$ and $e_k$ are respectively the volume 
fraction, the density, the pressure and the internal energy of the 
\emph{k-th}\,($1$ or $2$) component of the flow. The volume fractions for 
both fluids have to satisfy the saturation restriction $\sum\alpha_k = 1$ 
and the interfacial pressure $P_i$ is defined as 
$P_i= \alpha_1p_1+ \alpha_2p_2$. Additionally, the mixture density $\rho$, 
velocity $u$, pressure $p$ and internal energy $e$ are defined as:
$$
\rho = \alpha_1\rho_1+ \alpha_2\rho_2,
$$
$$
u = (\alpha_1\rho_1 u_1 + \alpha_2\rho_2 u_2)/\rho,
$$
$$
p = \alpha_1 p_1 +\alpha_2 p_2,
$$
$$
e = (\alpha_1\rho_1 e_1 + \alpha_2\rho_2 e_2)/\rho.
$$
The $\mu$ variable represents a homogenization parameter controlling 
the rate at which pressure tends towards equilibrium and it depends
on the compressibility of each fluids and their interface topology. 
Its physical meaning was justified using the second law of 
thermodynamics~\cite{Baer1986}. 
Instead of using only mixture thermodynamic variables, the model~(\ref{eq:1}) 
keeps two distinct pressures. As a result, the thermodynamic non-equilibrium 
source term $\mu(p_1-p_2)$ exists in the volume fraction evolution equation 
and the source term $P_i\mu(p_1-p_2)$ in the energy conservation equations.

On the one hand the presence of the left hand side non-conservative terms 
$\alpha_k p_k {\partial u}/{\partial x}$ complicates the 
analytical and computational treatment of the model~(\ref{eq:1}). 
The non-conservative terms do not allow the governing equations to be written 
in a divergence form, which is preferred for numerical handling of 
problems involving shocks. The classical Rankine-Hugoniot relations cannot be
defined in an unambiguous manner and additional relations or regularisation 
procedures must be proposed instead.

On the other hand when dealing with the bubble interface 
represented by the density jump, these non-conservative terms enable 
accommodating thermodynamic non-equilibrium effects between the bubble 
and its surrounding during the passage of shock waves. 
The pressure non-equilibrium state can be solved using the instantaneous 
relaxation model with the efficient numerical algorithm proposed 
in~\cite{Saurel2009}. This model, while retaining the separate equations 
of state and pertinent energy equation on both sides of the interfaces, 
introduces an additional total mixture energy equation.  As a result shock 
waves can be correctly transmitted through the heterogeneous media and the 
volume fraction positivity in the numerical solution is preserved. This key 
equation in~\cite{Saurel2009} was derived by combining the two internal 
energy equations with mass and momentum equations. The final form of the 
total mixture energy equation can be written as: 
\begin{equation}\label{eq:2}
\pd{\rho(Y_1e_1+Y_2e_2+\frac{1}{2}u^2)}{t}\\+ \pd{u(\rho(Y_1e_1+Y_2e_2+\frac{1}{2}u^2)+(\alpha_1p_1+\alpha_2p_2))}{x} = 0,
\end{equation}
where, $Y_1$ and $Y_2$ are the mass fractions with general form 
$Y_k=\alpha_k\rho_k/\rho$.
The numerical procedures discussed in the next section tackle 
the overdetermined system of equations consisting 
of ~(\ref{eq:1}) and~(\ref{eq:2}) and correct the errors resulting 
from the numerical integration of the non-conservative terms: 
$\alpha_k p_k {\partial u}/{\partial x}$.
The solution aspects of the overdetermined hyperbolic systems have been 
considered earlier (see e.g.~\cite{Godunov2008}).

The mixture sound speed in this six-equation model has the desired 
monotonic behaviour as a function of volume and mass fractions and 
is expressed as:
$$
c^2=Y_1c_1^2 + Y_2c_2^2,
$$
where, $c_1$ and $c_2$ are the speeds of sound  of the pure fluids.

The model~(\ref{eq:1}) is supplemented by a thermodynamic closure. The 
ideal gas equation of state, relating the internal energy to the pressure 
$p=p(\rho,e)$, is used for both flow components experiencing different 
thermodynamic states. 
For a given fluid, the equation of state can be written as a pressure law:
\begin{equation}
p_k = (\gamma_k-1)\rho_k e_k,
\label{eq_state}
\end{equation}
where $\gamma_k$ is the specific heat ratio for the component $k$ of the 
flow. Similarly, the mixture equation of state takes the form:
\begin{equation}\label{meq_state}
p = (\gamma-1)\rho e,
\end{equation}
where the specific heat ratio of mixed gas $\gamma$ is calculated from:
$$
\frac{1}{\gamma-1} = \sum_k \frac{\alpha_k}{\gamma_k-1}.
$$

In this two-component formulation there are no explicit diffusive terms. 
These terms can be neglected as the diffusive effects do not play a major
role in the early stages of bubble-shock interaction. The calculation 
of the kinematic viscosity of the mixture and estimation of resulting 
viscous length scales were provided in~\cite{Picone1988,Schilling2007}. 
The viscosities of the considered fluids are ($\mu \sim 10^{-5}~\rm{Pa.s}$)
and the evolution is studied over short time scales 
($t \leqslant 10^{-3}~\rm{s}$).

To tackle two dimensional geometries and two component flows, the original 
six-equation model~(\ref{eq:1}) can be extended to the following form:
\begin{eqnarray}\label{sy:2deq}
&&\pd{\alpha_1}{t}+u\pd{\alpha_1}{x} +v\pd{\alpha_1}{y}= \mu(p_1-p_2),\nonumber\\
&&\pd{\alpha_1\rho_1}{t}+\pd{\alpha_1\rho_1u}{x}+\pd{\alpha_1\rho_1v}{y} = 0,\nonumber\\
&&\pd{\alpha_2\rho_2}{t}+\pd{\alpha_2\rho_2u}{x}+\pd{\alpha_2\rho_2v}{y} = 0,\nonumber\\
&&\pd{\rho u}{t}+\pd{[\rho u^2+(\alpha_1p_1+\alpha_2p_2)]}{x} +\pd{\rho u v}{y}= 0,\\
&&\pd{\rho v}{t}+\pd{\rho u v}{x}+\pd{[\rho v^2+(\alpha_1p_1+\alpha_2p_2)]}{y} = 0,\nonumber\\
&&\pd{\alpha_1\rho_1 e_1}{t}+\pd{\alpha_1\rho_1 e_1u}{x}+\pd{\alpha_1\rho_1 e_1v}{y}+\alpha_1p_1\pd{u}{x}+\alpha_1p_1\pd{v}{y} = P_i \mu(p_1-p_2),\nonumber\\
&&\pd{\alpha_2\rho_2 e_2}{t}+\pd{\alpha_2\rho_2 e_2u}{x}+\pd{\alpha_2\rho_2 e_2v}{y}+\alpha_2p_2\pd{u}{x}+\alpha_2p_2\pd{v}{y} = -P_i \mu(p_1-p_2),\nonumber
\end{eqnarray}
and
\begin{equation}\label{eq:6}
\pd{\rho E}{t}+\pd{u(\rho E+(\alpha_1p_1+\alpha_2p_2))}{x}+\pd{v(\rho E+(\alpha_1p_1+\alpha_2p_2))}{y} = 0.
\end{equation}
where, $u$ and $v$ represent the components of the velocity
in the $x$ and $y$ directions, respectively. The total energy 
of the mixture for two dimensional flows is given by 
$E = Y_1e_1+Y_2e_2+\frac{1}{2}u^2+\frac{1}{2}v^2$.

\section{\label{sec:3}Numerical method} 
As stated in the previous section, model~(\ref{sy:2deq}) cannot be written 
in a divergence form and hence the standard numerical methods developed for 
conservation laws are not applied directly here. In order to solve this system 
a numerical scheme is constructed that decouples the left hand side of 
model~(\ref{sy:2deq}) from the pressure relaxation source terms 
(the right hand side of the model). The left hand side, which represents 
the advection part of the flow equations, is then analysed to determine 
the mathematical structure of the system and is rewritten in terms 
of the primitive variables as follows:
\begin{equation}
\pd{W}{t}+A(W)\pd{W}{x}+B(W)\pd{W}{y} = 0,
\label{eqn:7}
\end{equation}
where the vector of primitive variables $W$, Jacobian matrices $A(W)$ 
and $B(W)$ for the extended model~(\ref{sy:2deq}) are:
$$
W = \left[\begin{matrix}\alpha_1\\
                                          \rho_1\\
                                          \rho_2\\
                                           u\\
                                           v\\
                                         p_1\\
                                         p_2\end{matrix}\right]
,
A(W) = \left[\begin{matrix}u & 0 & 0 & 0 & 0 & 0 & 0\\
                           0 & u & 0 & \rho_1 & 0 & 0 & 0\\
                           0 & 0 & u & \rho_2 & 0 & 0 & 0\\
                           \frac{p_1-p_2}{\rho} & 0 & 0 & u & 0 &\frac{\alpha_1}{\rho} & \frac{1-\alpha_1}{\rho}\\
                            0 & 0 & 0 & 0 & u & 0 & 0\\
                            0 & 0 & 0 & \rho_1 c_1^{2} & 0 & u & 0\\ 
                           0 & 0 & 0 & \rho_2 c_2^{2} & 0 & 0 & u
\end{matrix}\right]
$$
and
$$
B(W) = \left[\begin{matrix}v & 0 & 0 & 0 & 0 & 0 & 0\\
                           0 & v & 0 & 0 & \rho_1  & 0 & 0\\
                           0 & 0 & v & 0 & \rho_2 & 0 & 0\\
                           0 & 0 & 0 & v & 0 & 0 & 0 &\\
                          \frac{p_1-p_2}{\rho} & 0 & 0 & 0 & v &\frac{\alpha_1}{\rho} & \frac{1-\alpha_1}{\rho}\\
                           0 & 0 & 0 & 0 & \rho_1 c_1^{2} & v & 0\\
                           0 & 0 & 0 & 0 & \rho_2c_2^{2} & 0 & v
\end{matrix}\right]
$$
The seven eigenvalues of the Jacobian matrix $A(W)$ are determined to be: 
$u-c$, $u$, $u$, $u$, $u$, $u$ and $u+c$. Similarly, the eigenvalues of 
the Jacobian matrix $B(W)$ are: $v-c$, $v$, $v$, $v$, $v$, $v$ and $v+c$.

\subsection{Solution of the hyperbolic part}
The above primitive form~(\ref{eqn:7}) is hyperbolic but not strictly 
hyperbolic. Indeed some eigenvalues, which represent the wave speeds, 
are real but not distinct. The solution of this hyperbolic problem is 
obtained using the extended Godunov scheme. To achieve second order 
accuracy the numerical algorithm is equipped with the classical Monotonic
Upstream-centered Scheme for Conservation Laws (MUSCL)~\cite{Toro1999}.

\begin{figure}
\includegraphics [width=0.4\textwidth]{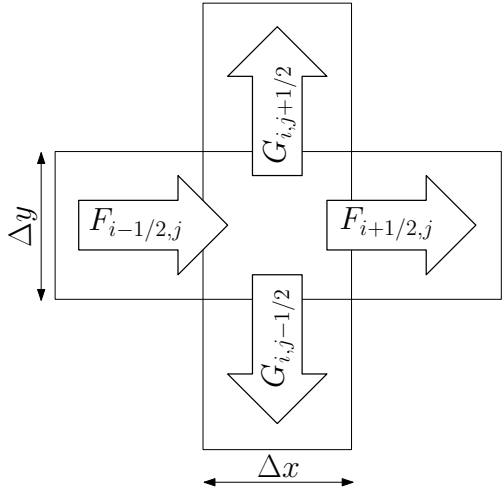}
\caption{Numerical flux notations within a 2D quadratic mesh.}
\label{Fig1}
\end{figure}

\begin{figure}
\includegraphics[width=0.5\textwidth]{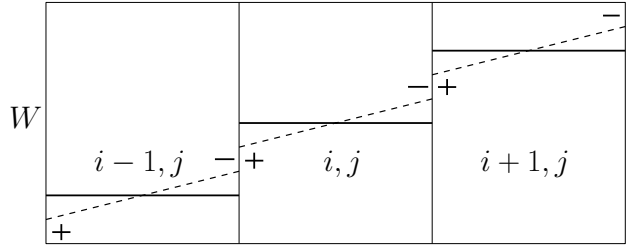}
\caption{Schematic diagram representing notations at each computational cell for the piecewise linear variable states (dash line) in the second order scheme.}
\label{Fig2}
\end{figure}

The splitting scheme has been applied to solve the conservative equations 
on regular meshes. In each time step of simulation ($\Delta t$), 
the conservative variables evolve in alternate directions $(x,y)$ during time
sub-steps ($\Delta t/2$) which are denoted by ($U^{n+1/2}$) and ($U^{n+1}$). 
The time increment for the 2D second order discretisation of the Godunov 
scheme takes the following sub-steps: 
\begin{equation*} 
   U_{i,j}^{n+1/2}=U_{i,j}^n-\frac{\Delta t}{\Delta x}[F^{n}(U^{*}(U_{i+\frac{1}{2},j}^{-}, U_{i+\frac{1}{2},j}^{+}))\\ -  F^{n}(U^{*}(U_{i-\frac{1}{2},j}^{-}, U_{i-\frac{1}{2},j}^{+}))]
\end{equation*}
 and
 \begin{equation*} 
U_{i,j}^{n+1}=U_{i,j}^{n+1/2} -\frac{\Delta t}{\Delta y}[G^{n+1/2}(U^{*}(U_{i,j+\frac{1}{2}}^{-}, U_{i,j+\frac{1}{2}}^{+}))\\ -  G^{n+1/2}(U^{*}(U_{i,j-\frac{1}{2}}^{-}, U_{i,j-\frac{1}{2}}^{+}))].
\end{equation*}
The components of the vector 
$U=[\alpha_1\rho_1, \alpha_2\rho_2, \rho u, \rho v, \rho E]^T$ 
are the conservative variables. $U_{i,j}^n$ represents the state vector 
in a cell ($i,j$) at time $n$ and $U_{i,j}^{n+1}$ represents the state 
vector at the next time step. $F(U^*)$ is the flux function 
in the $x$ direction
$F(U)=[\alpha_1\rho_1 u, \alpha_2\rho_2 u, \rho u^2 +p, \rho uv, u(\rho E+p)]^T$. $G(U^*)$
is the flux function in the $y$ direction 
$G(U)=[\alpha_1\rho_1 v, \alpha_2\rho_2 v, \rho uv, \rho v^2 +p, v(\rho E+p)]^T$. 
The superscript \textquotedblleft *\textquotedblright refers to the state
at each cell boundary. Figure~\ref{Fig1} presents a diagram for the flux 
configurations in two dimensional computations. The plus and minus signs 
refer to the conservative variable and flux values at cell boundaries in 
the second order scheme. The second order accuracy is achieved by applying 
three major steps: the first step consists in the reconstruction of 
the average local values in each computational cell using extrapolation 
of piecewise linear approximations, the second step consists in 
the determination of the variable values at a middle time step and finally 
the last step is the solution of the Riemann problem. Figure~\ref{Fig2} 
shows the piecewise linear variable variation at the boundaries of each 
cell. The flux functions in the Godunov scheme are obtained using 
the approximate Harten, Lax and Van Lear (HLL) Riemann 
solver~\cite{Harten1983,Toro1999}.

\begin{figure}
\includegraphics [width=0.4\textwidth]{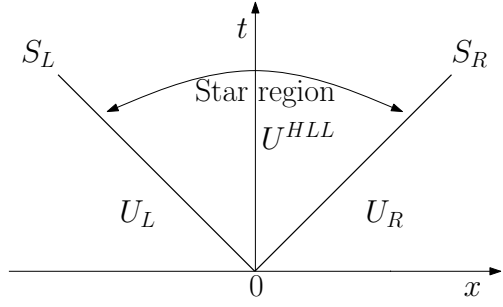}
\caption{Wave configuration at a cell boundary in the approximate HLL Riemann solver 
with right initial data $(U_R)$, left initial data $(U_L)$ and the star region.}
\label{Fig3}
\end{figure}

Figure~\ref{Fig3} shows a schematic diagram of the typical wave configuration
in the approximate HLL Riemann solver. The wave speeds $S^R$ and $S^L$ 
are the boundaries of three characteristics regions: left region ($U_L$), 
right region ($U_R$) and the star region ($U^{HLL}$). In this Riemann solver,
the second order numerical flux functions in the star region at each cell 
boundary are written as:

 \[ F_{i+\frac{1}{2},j} = \left\{ \begin{array}{ll}
F_{i,j}                                            & \mbox{if $0 \le S_{i+\frac{1}{2},j}^L$},\\
\frac{S_{i+\frac{1}{2},j}^{R}F_{i,j} - S_{i+\frac{1}{2},j}^{L}F_{i+1,j}+S_{i+\frac{1}{2},j}^{R}S_{i+\frac{1}{2},j}^{L}(U_{i+1,j}-U_{i,j})}{S_{i+\frac{1}{2},j}^{R}-S_{i+\frac{1}{2},j}^{L}}  & \mbox{if $S_{i+\frac{1}{2},j}^L \le 0 \le S_{i+\frac{1}{2},j}^R$},\\
F_{i+1,j}                                            & \mbox{if $0 \ge S_{i+\frac{1}{2},j}^R$}.
          \end{array} \right.\]
          
\[ F_{i-\frac{1}{2},j} = \left\{ \begin{array}{ll}
F_{i-1,j}                                            & \mbox{if $0 \le S_{i-\frac{1}{2},j}^L$},\\
\frac{S_{i-\frac{1}{2},j}^{R}F_{i-1,j} - S_{i-\frac{1}{2},j}^{L}F_{i,j}+S_{i-\frac{1}{2},j}^{R}S_{i-\frac{1}{2},j}^{L}(U_{i,j}-U_{i-1,j})}{S_{i-\frac{1}{2},j}^{R}-S_{i-\frac{1}{2},j}^{L}}  & \mbox{if $S_{i-\frac{1}{2},j}^L \le 0 \le S_{i-\frac{1}{2},j}^R$},\\
F_{i,j}                                            & \mbox{if $0 \ge S_{i-\frac{1}{2},j}^R$}.
          \end{array} \right.\]
and
 \[ G_{i,j+\frac{1}{2}} = \left\{ \begin{array}{ll}
G_{i,j}                                            & \mbox{if $0 \le S_{i,j+\frac{1}{2}}^L$},\\
\frac{S_{i,j+\frac{1}{2}}^{R}G_{i,j} - S_{i,j+\frac{1}{2}}^{L}G_{i,j+1}+S_{i,j+\frac{1}{2}}^{R}S_{i,j+\frac{1}{2}}^{L}(U_{i,j+1}-U_{i,j})}{S_{i,j+\frac{1}{2}}^{R}-S_{i,j+\frac{1}{2}}^{L}}  & \mbox{if $S_{i,j+\frac{1}{2}}^L \le 0 \le S_{i,j+\frac{1}{2}}^R$},\\
G_{i,j+1}                                            & \mbox{if $0 \ge S_{i,j+\frac{1}{2}}^R$}.
          \end{array} \right. \]
          
\[ G_{i,j-\frac{1}{2}} = \left\{ \begin{array}{ll}
G_{i,j-1}                                            & \mbox{if $0 \le S_{i,j-\frac{1}{2}}^L$},\\
\frac{S_{i,j-\frac{1}{2}}^{R}G_{i,j-1} - S_{i,j-\frac{1}{2}}^{L}G_{i,j}+S_{i,j-\frac{1}{2}}^{R}S_{i,j-\frac{1}{2}}^{L}(U_{i,j}-U_{i,j-1})}{S_{i,j-\frac{1}{2}}^{R}-S_{i,j-\frac{1}{2}}^{L}}  & \mbox{if $S_{i,j-\frac{1}{2}}^L \le 0 \le S_{i,j-\frac{1}{2}}^R$},\\
G_{i,j}                                            & \mbox{if $0 \ge S_{i,j-\frac{1}{2}}^R$}.
          \end{array} \right. \]
Similarly, the splitting scheme is applied in the descritization of 
non-conservative equations, i.e. the volume fraction and the two energy 
equations as follows:
\begin{equation*}
\alpha_{1i,j}^{n+1/2}=\alpha_{1i,j}^n-\frac{\Delta t}{\Delta x}[(u\alpha_1)_{i+\frac{1}{2},j}^{*} -(u\alpha_1)_{i-\frac{1}{2},j}^{*}\\-\alpha_{1i,j}( u_{i+\frac{1}{2},j}^*-u_{i-\frac{1}{2},j}^*)]^n,
\end{equation*}
\begin{equation*}
\alpha_{1i,j}^{n+1}=\alpha^{n+1/2}_{1i,j}-\frac{\Delta t}{\Delta y}[(v\alpha_1)_{i,j+\frac{1}{2}}^{*} -(v\alpha_1)_{i,j-\frac{1}{2}}^{*}\\-\alpha_{1i,j}( v_{i,j+\frac{1}{2}}^*-v_{i,j-\frac{1}{2}}^*)]^{n+1/2},
\end{equation*}
\begin{equation*}
(\alpha\rho e )_{ki,j}^{n+1/2}=(\alpha\rho e )_{ki,j}^n-\frac{\Delta t}{\Delta x}[(\alpha\rho eu)_{ki+\frac{1}{2},j}^{*} -(\alpha\rho eu)_{ki-\frac{1}{2},j}^{*}\\+(\alpha p)_{ki,j}( u_{i+\frac{1}{2},j}^*-u_{i-\frac{1}{2},j}^*)]^n,
\end{equation*}
\begin{equation*}
(\alpha\rho e )_{ki,j}^{n+1}=(\alpha\rho e )_{ki,j}^{n+1/2}-\frac{\Delta t}{\Delta y}[(\alpha\rho ev)_{ki,j+\frac{1}{2}}^{*} -(\alpha\rho ev)_{ki,j-\frac{1}{2}}^{*}\\+(\alpha p)_{ki,j}( v_{i,j+\frac{1}{2}}^*-v_{i,j-\frac{1}{2}}^*)]^{n+1/2},
\end{equation*}       
The stability of the numerical method is controlled by 
the Courant-Friedrichs-Lewy (CFL) number, which imposes a restriction on 
the time step $\Delta t$ as follows:
\begin{equation}
\Delta t = \text{CFL} \times\text{min}(\frac{\Delta x}{S_x},\frac{\Delta y}{S_y}),
\end{equation}
where $S_x$ and $S_y$ are the maximum wave speeds in the $x$ and $y$ 
directions respectively.
\begin{equation*}
    S_x = \text{max} \langle 0, u_{i\pm\frac{1}{2},j}^{+}+ c_{i\pm\frac{1}{2},j}^{+}, u_{i\pm\frac{1}{2},j}^{-}+ c_{i\pm\frac{1}{2},j}^{-}\rangle,
   \end{equation*}
   \begin{equation*}
    S_y = \text{max} \langle 0, v_{i,j\pm\frac{1}{2}}^{+}+c_{i,j\pm\frac{1}{2}}^{+}, v_{i,j\pm\frac{1}{2}}^{-}+c_{i,j\pm\frac{1}{2}}^{-}\rangle.
   \end{equation*}
\subsection{Solution of the pressure relaxation part}\label{pr:relax}
In each time step after the hyperbolic advection part is accomplished, 
the pressure equilibrium is achieved via the relaxation procedure. 
The pressure relaxation implies volume variations because of 
the interfacial pressure work.
This represents the solution of the sub-problem governed by the following 
ordinary differential equations (ODE), with the source term representing 
the right hand side of the model~(\ref{sy:2deq}). 
\begin{eqnarray}
\frac{\partial\alpha_1}{\partial t} &= & \mu (p_1 - p_2),\nonumber\\
\frac{\partial\alpha_1\rho_1}{\partial t} &=& 0, \nonumber\\
\frac{\partial\alpha_2\rho_2 }{\partial t} &=& 0,  \nonumber \\
\frac{\partial\rho u}{\partial t} &=& 0, \nonumber\\
\frac{\partial\rho v}{\partial t} &=& 0, \nonumber\\
\frac{\partial\alpha_1\rho_1e_1}{\partial t} &=& \mu P_i (p_1 - p_2),\nonumber\\
\frac{\partial\alpha_2\rho_2e_2}{\partial t} &=&  -\mu P_i (p_1 - p_2).
\label{pressure relx}
\end{eqnarray}
The pressure relaxation is fulfilled instantaneously when the value 
of $\mu$ in system~(\ref{sy:2deq}) is assumed to be infinite. To solve 
the ODE system~(\ref{pressure relx}), an iterative method for the pressure 
relaxation for compressible multiphase flow is implemented. This method 
is the iterative \emph{\textquotedblleft procedure 4\textquotedblright} 
described in~\cite{Lallemand2005}. This step rectifies the calculation 
of the internal energies to satisfy the second law of thermodynamics. 
The second amendment comes from solving the extra total energy 
equation~(\ref{eq:6}), where the mixture pressure is calculated 
from the mixture equation of state. 
The overall sequence of the numerical solution steps follows the 
idea of succession of operators introduced in~\cite{Strang1968}. 
\section{\label{sec:4}Results and Discussion}           
This section is divided into four parts: In the first part the general 
description of the physics and the mechanism of the shock-bubble interaction
phenomenon are revisited. In the second part the correctness of the results 
obtained using the developed numerical code is quantified. This is made by 
validating the numerical results for different shock-bubble interaction 
scenarios against the experimental data reported in~\cite{Layes2007}. 
In the third and fourth subsections the investigation of the shock-bubble 
interaction problem is extended to a wider selection of physically 
intriguing cases for which experimental data are not available.
\subsection{General features of shock-bubble interaction problems}
The flow configurations for the studied problems are classified according 
to the value of the Atwood number $A = (\rho_b - \rho_s) / (\rho_b + \rho_s)$,
where $\rho_b$ is the density of the bubble and $\rho_s$ is the density 
of the surrounding medium. If the density of the bubble is lower than 
the density of the surrounding fluid, the value of the Atwood number 
becomes negative and this case represents heavy/light arrangement. 
In contrast, if the density of the bubble is higher than the density 
of the surrounding fluid, the value of the Atwood number becomes positive 
and this case represents light/heavy arrangement. Alternative terminologies 
exist and describe these flow configurations as divergent in the case of 
light bubble or as convergent in the case of heavy bubble~\cite{Haas1987},
or fast/slow interface or slow/fast interface according to the sound speed 
of the flow constituents~\cite{Zabusky1998}.    

\begin{figure}
\includegraphics[width=0.53\textwidth]{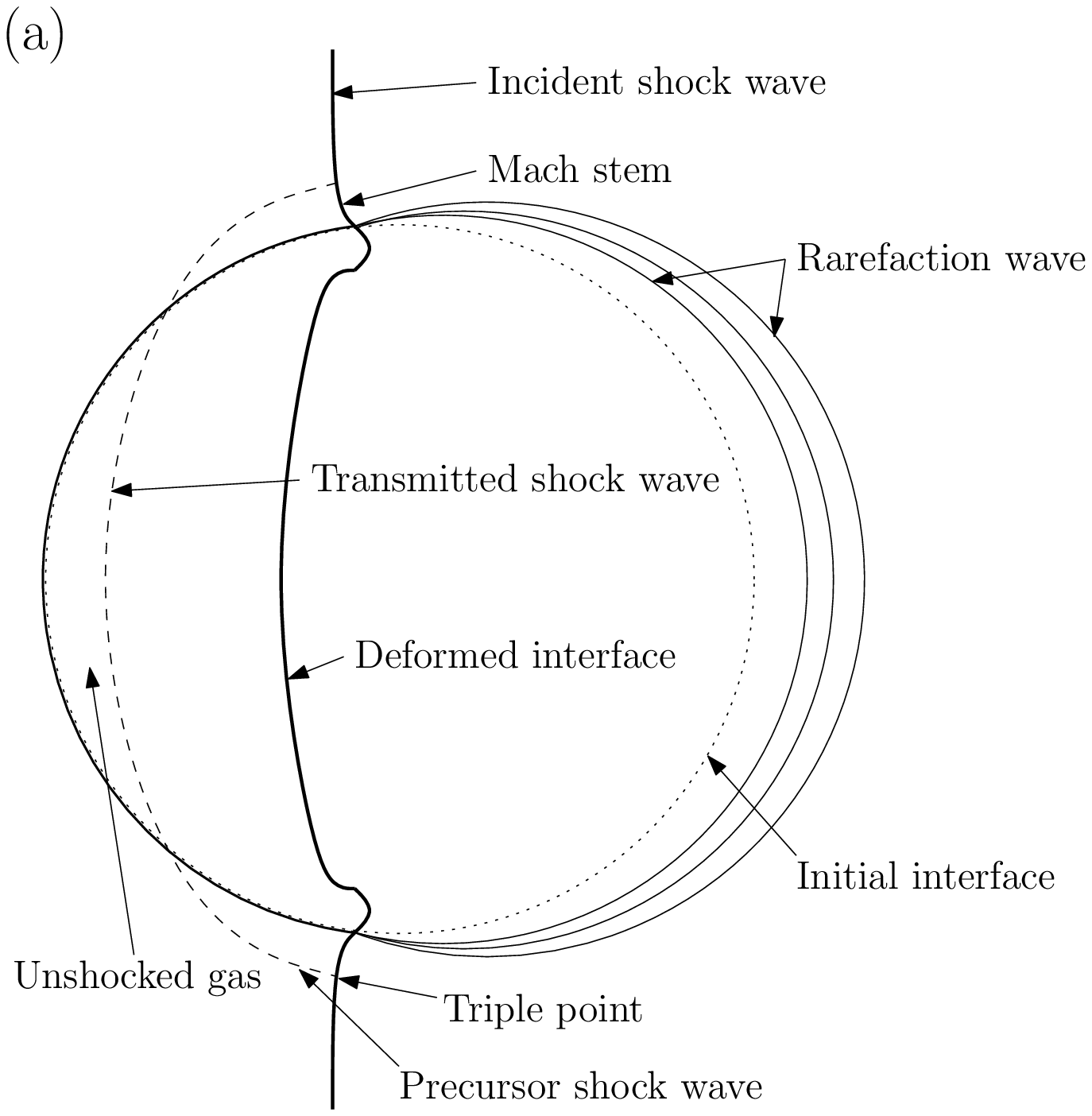}
\includegraphics[width=0.53\textwidth]{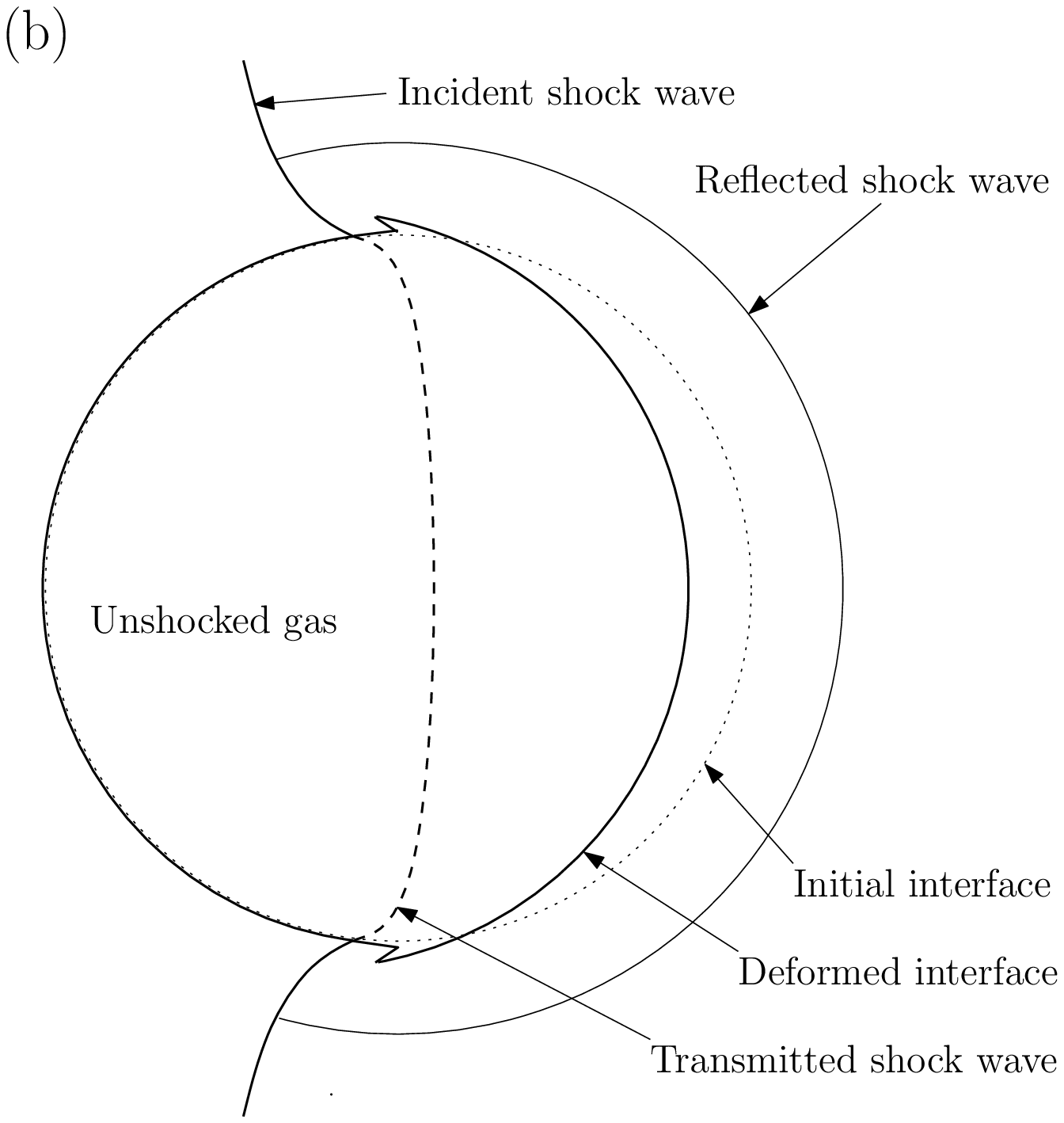}
\caption{Schematic diagram of a shock-bubble interaction flow field and 
different wave configurations (a) light bubble and (b) heavy bubble. 
The  planar shock moves from right to left.}
\label{Fig4}
\end{figure}

Figure~\ref{Fig4} schematically presents the typical flow configuration 
during an early stage of the shock-bubble interaction process. The flow patterns
of heavy/light, Fig.~\ref{Fig4}(a), and light/heavy, Fig.~\ref{Fig4}(b), 
scenarios show the set of wave configurations associated with the interaction
and the deformation of the bubble interface. After hitting the upstream 
interface of the bubble from the right hand side, the planar shock wave changes
its uniform front, which evolves into two parts. One part does not interact 
directly with the gas filling the bubble while the second one transforms into
a transmitted wave interacting with the bubble. In addition a reflected wave, 
as in the case of a heavy bubble, or a rarefaction wave front, as in the case
of a light bubble, does occur and propagates back in the right direction.

In the case of a light bubble (i.e. negative $A$ number, Fig.~\ref{Fig4}(a)), 
the transmitted shock travels through the gas bubble faster than the incident 
shock outside the bubble. This is the consequence of the mismatch in flow 
constituents acoustic impedances ($Z = \rho c$) across the interface. 
The shock front also takes a divergent shape due to the curvature of the 
interface and provokes the generation of a set of secondary waves inside 
and outside the bubble boundary. These secondary waves consist in irregular 
waves~\cite{Henderson1966, Henderson1989}. A precursor (refracted) shock 
wave propagates downstream outside the bubble, internal reflected shocks 
are generated inside the bubble and move back upstream as the result 
of the interaction of the transmitted shock with the internal surface 
of the bubble. A Mach stem shock wave travels outside the bubble. 
A triple point is formed outside the bubble owing to the intersection 
of the exterior incident shock, the precursor shock and the Mach stem. 

In the case of a heavy bubble (i.e. positive $A$ number, Fig.~\ref{Fig4}(b))
the scenario is completely different. The difference in the acoustic impedance
between the fluids across the interface makes the transmitted shock inside the
bubble moving more slowly than the incident shock outside the bubble. 
The transmitted shock becomes convergent. The interaction of the transmitted 
shock with the internal surface of the bubble produces a rarefaction wave 
propagating backward inside the bubble. 

To explain the role of acoustic impedance mismatch in the creation 
of a vorticity field it is convenient to consider the vorticity 
transport form of the Euler equations. The momentum equation governing 
the evolution of vorticity is 

\begin{equation}\label{eq:omega}
\frac{D\omega}{D t} = 
\underbrace{\vphantom{\frac{1}{\rho^2}}(\omega \cdot \nabla)\bf{u}}_\text{stretching}
-\underbrace{\vphantom{\frac{1}{\rho^2}}\omega(\nabla \cdot {\bf{u}})}_\text{dilatation}
+\underbrace{\frac{1}{\rho^2}(\nabla\rho \times \nabla p)}_\text{baroclinicity}.
\end{equation}

This equation contains, in contrast to its 2D counterpart, the term 
corresponding to vorticity stretching. High resolution 
shock-induced 3D simulations were performed to analyze 
the relative importance of stretching, dilatation and baroclinic terms 
in the vorticity equation at 
$\textit{Ma}=3$ and $\textit{Ma}=10$~\cite{Hejazialhosseini2013}. 
It was found that the stretching term contribution manifests its existence  
after initial phases of shock bubble interaction.
The only term of importance in the early stages of a shock-bubble interaction
for which $\omega$  is initially equal to zero is the baroclinic torque 
($\nabla\rho \times \nabla p$). The misalignment of the local pressure 
and density gradients leads to the non-zero source term 
in equation~(\ref{eq:omega}). Because of the curved surface of the bubble, 
different parts of the incident planar shock will strike the bubble surface
at different times, so the refracted interior wave will be misaligned with
the density gradient. The baroclinic torque is the largest where the pressure
gradient is perpendicular to the density gradient. Whereas, at the most 
upstream and downstream poles of the gas bubble, the baroclinic torque 
is equal to zero, owing to the collinearity of density and pressure gradients. 
The curvature of the shock wave front (the refracted shock wave) has
also been used to build a theory behind the vorticity generation. It originates 
from the conservation of the tangential velocity and the angular momentum across 
the shock wave as the compression only affects the motion normal to the shock 
surface (for details see \cite{Kevlahan1997} and also recent works of
~\cite{Huete2011, Huete2012, Wouchuk2015}). The rotational motion
starting from zero-vorticity initial conditions distorts the flow field
and the shape of the bubble.  The lighter density fluid will be accelerated
faster than the high density fluid.

\subsection{Validation of shock-single bubble interaction cases}
The experimental studies performed in~\cite{Layes2007} and \cite{Zhai2011} 
constituted validation cases for the numerical approach developed 
in the present paper. The authors provided sequences of flow structures 
resulting from experiments and compared them with numerical simulations 
they also conducted. The mathematical model and numerical technique 
considered in the present contribution are fundamentally different 
from the ones utilised in the reference works  
of~\cite{Layes2007} and \cite{Zhai2011} and hence the main features 
of their approaches are highlighted.

The authors in~\cite{Layes2007} employed a homogenization
method known as the discrete equations method (DEM), which was earlier 
introduced by~\cite{Abgrall2003}. In this approach the averaged equations 
for the mixture are not used. Instead the DEM method obtains a well-posed 
discrete equation system from the single-phase Euler equations by 
construction of a numerical scheme which uses a sequence of accurate 
single-phase Riemann solutions. The local interface variables 
are determined at each two-phase interface. Then, an averaging procedure 
which enables coupling between the two fluids is applied generating 
a set of discrete equations that can be used directly as a numerical scheme. 
The advantage of such an approach is its natural ability to treat correctly
the non-conservative terms. In our approach the solution strategy to handle 
non-conservative terms is different. It requires the usage of an additional 
conservative equation for the total mixture energy (\ref{eq:2}). As a result 
the present model enables a correct transmission of shock waves through 
the heterogeneous media. The volume fraction positivity in the numerical 
solution is also preserved.

The authors in~\cite{Zhai2011} adopted the 2D axisymmetrical numerical 
approach of~\cite{Sun1999} and solved the mixture Euler equations 
supplemented by one species conservation equation to capture the interface. 
It is assumed in this approach that the gas components are in pressure
equilibrium and move with a single velocity. This assumption restricts 
the approach to the cases when the density variations between components 
are moderate.   

%%%%%%%%%%%%%%%%%%
\subsubsection{Experiments of Layes and Le\,M\'{e}tayer~\cite{Layes2007}}
\label{Layes_validation}
\begin{figure}
\includegraphics[width=0.9\textwidth]{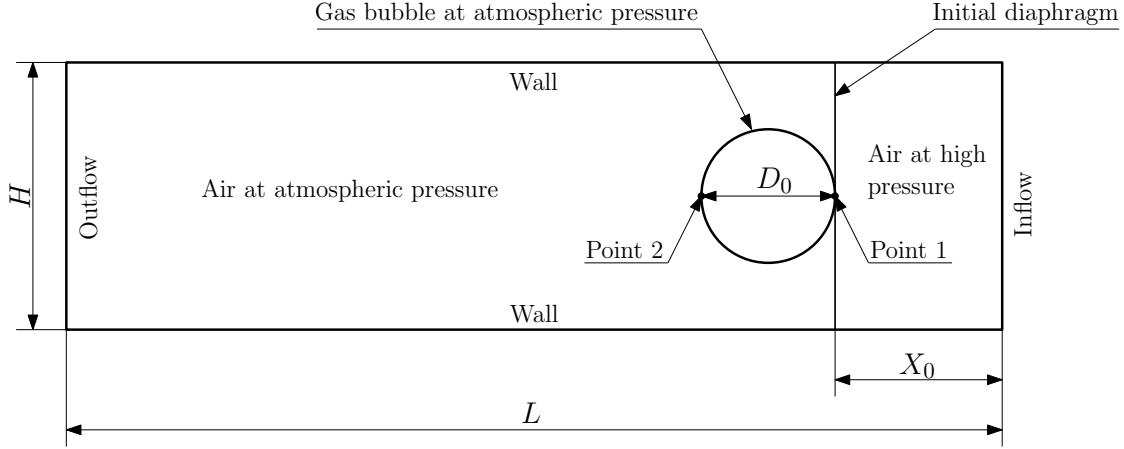}\\
\caption{Schematic diagram of the initial state of shock-bubble interaction 
test problems.}
\label{Fig5}
\end{figure}

%%%%%%%%%%%%%%%%%%
These experiments were reproduced numerically for three different cases 
in which $\textit{Ma} = 1.5$ planar shock wave interacts with a helium, nitrogen 
or krypton cylindrical bubble of a diameter $D_o = 0.04~\rm{m}$ located 
in a shock tube. The thermodynamic properties of the bubbles and surrounding 
air are given in Table~\ref{Table1}. The schematic diagram of the computational 
domain and the initial set-up is shown in Fig.~\ref{Fig5}. The shock tube 
dimensions are $L = 0.3~\rm{m}$ and $H = 0.08~\rm{m}$. The initial position 
of the shock is $X_o = 0.05~\rm{m}$. The solid walls are treated as reflecting 
boundary conditions. The inflow boundary conditions are set to the exact 
pre-shock region parameters summarised in Table~\ref{Table2} and the standard 
zero-order extrapolation is used as the outflow boundary conditions. 
In all three cases the bubble was initially assumed to be in mechanical and 
thermal equilibrium with the surrounding air. The shock wave propagates
in the air from right to left and impacts the bubble. The Atwood numbers are 
listed in Table~\ref{Table3} and represent three distinct regimes 
of shock-bubble interactions.  

\begin{table}
\caption{\label{Table1}Properties of air and different gas bubbles at standard conditions.}
\begin{ruledtabular}
\begin{tabular}{lcccc}
Physical property & Helium & Air & Nitrogen & Krypton\\
\hline
Density, $kg/m^3$  & 0.167 & 1.29 & 1.25 & 3.506\\
Sound speed, $m/s$ &1007 & 340 & 367 &220\\
Heat capacity ratio $\gamma$ & 1.67 & 1.4 & 1.67 & 1.67\\
Acoustic impedance, $Pa\cdot s/m$ & 168.16 & 421.25 & 428.28 & 771.32
\end{tabular}
\end{ruledtabular}
\end{table}

\begin{table}
\caption{\label{Table2}Properties of air at the high pressure chamber.}
\begin{ruledtabular}
\begin{tabular}{ll}
Property   & Value \\
\hline
Density, $kg/m^3$ & 2.4021\\
Pressure, $Pa$ & $2.4909\times10^5$\\
Shock Mach number &1.5\\
\end{tabular}
\end{ruledtabular}
\end{table}

\begin{table}
\caption{\label{Table3}Atwood number for air-bubble configurations.}
\begin{ruledtabular}
\begin{tabular}{ll}
Air/bubble configuration&   Atwood number\\         
\hline
Air/Helium&  $-0.7708$\\   
Air/Nitrogen& $-0.0157$ \\
Air/Krypton &  $+0.4621$  \\
\end{tabular}
\end{ruledtabular}
\end{table}

The first case corresponds to the interaction of a shock wave with 
a helium bubble surrounded by ambient air. As the density of helium 
is lower than the density of air this case represents a heavy/light 
interaction. The second test considers the interaction of the shock 
wave with a nitrogen bubble. Owing to the very small density ratio between 
nitrogen and air, this case is treated as an equal density problem. 
Finally the third case with a krypton bubble, which is heavier than air, 
represents a light/heavy interaction problem. 
The domain is discretised using a regular Cartesian grid consisting 
of $2700\times720$ cells, which corresponds to the resolution of $360$ cells 
across the bubble diameter. The CFL number is $0.3$. The present resolution
has been chosen based on information from numerical tests on meshes with 
different levels of refinement. Table~\ref{Table4} summarises 
the computational times for a selection of mesh resolutions and provides
the circulation values for air/He and air/Kr pairings at the physical 
time $60$~$\mu\rm{s}$.
The computational times are normalised by the longest simulation run. 
The difference between the total circulation values of the coarse mesh 
of $900\times240$ and the refined mesh of $2700\times720$ is only $1.5\%$.
Figures~\ref{Fig6} and~\ref{Fig7} show the convergence of the solution 
as the effective resolution is increased, by comparing the evolution of 
the pressure and density along the centre line of the domain 
($y = 0.04~\rm{m}$) at time $410$~$\mu\rm{s}$ for the air/He and air/Kr 
arrangements respectively. The changes in the thermodynamic values
are very small, especially between the two refined meshes. 

\begin{table}
\caption{\label{Table4}{Comparison of normalised computing 
times and total circulation {$\Gamma$(m$^2$/s)} values for 
selected simulation resolutions at time $t=60$~$\mu\rm{s}$.}}
\begin{ruledtabular}
\begin{tabular}{rlll}
mesh resolution & computational time & $\Gamma$ for air/He & $\Gamma$ for air/Kr\\
\hline
$900\times240$  & 0.20 & 7.968 & $-3.844$ \\
$1800\times480$ & 0.35 & 7.975 & $-3.890$ \\
$2700\times720$ & 1.00 & 7.972 & $-3.902$ \\
\end{tabular}
\end{ruledtabular}
\end{table}

\begin{figure}
\includegraphics[width=0.51\textwidth]{Fig6a.eps}
\includegraphics[width=0.48\textwidth]{Fig6b.eps}\\
\caption{Pressure,~(a) and density,~(b) distributions for shock-He bubble interaction along the tube centre line 
at $t= 410$~$\mu\rm{s}$ for different mesh resolutions.}
\label{Fig6}
\end{figure}

\begin{figure}
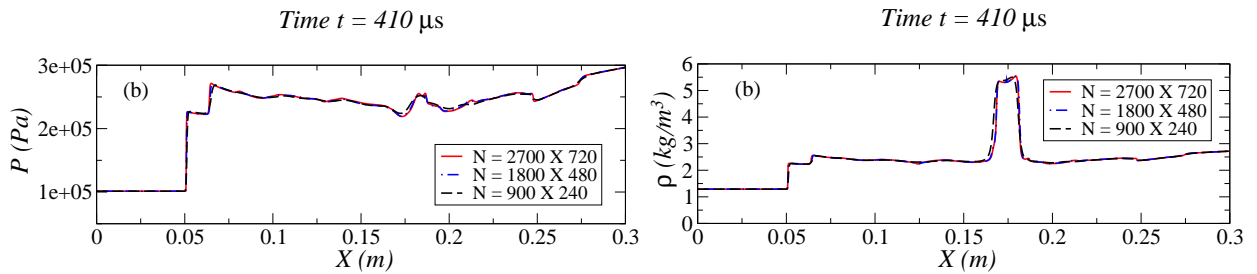

\includegraphics[width=0.51\textwidth]{Fig7a.eps}
\includegraphics[width=0.48\textwidth]{Fig7b.eps}\\
\caption{Pressure,~(a) and density,~(b) distributions for shock-Kr bubble interaction along the tube centre line 
at $t= 410$~$\mu\rm{s}$ for different mesh resolutions}
\label{Fig7}
\end{figure}

After performing numerical discretization tests, the experimental 
shadowgraph frames presented in~\cite{Layes2007}, were reproduced 
numerically. The computational simulations of the density field 
are presented by means of the idealized schlieren function at 
the same instants as in the experiment in~\cite{Layes2007}. Although the 
shapes of the deformed interfaces are recovered and can be observed 
clearly for different gas pairings it has to be noted that the accuracy 
of investigation is a function of experimental reproducibility. 
It is difficult to maintain the initial parameters of the shock 
wave Mach number, shape as well as size of the bubble and the gas 
composition from one to another experimental realisation~\cite{Layes2007}.
For example the tolerance for $\textit{Ma} = 1.5$ in the experiment was within
the range: $1.45-1.52$. In spite of these difficulties
the numerical results show a good approximation of the density contour 
plots obtained in the reference experiment. 

\begin{figure*}
\includegraphics[width=0.325\textwidth]{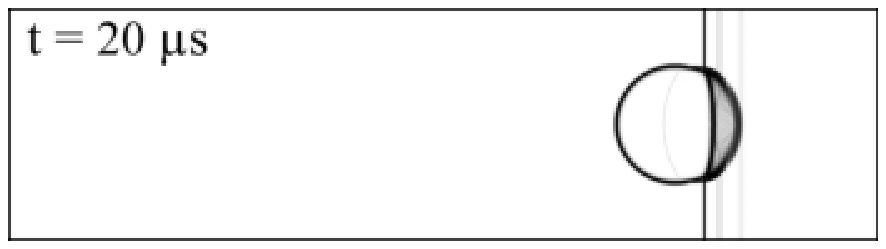}
\includegraphics[width=0.325\textwidth]{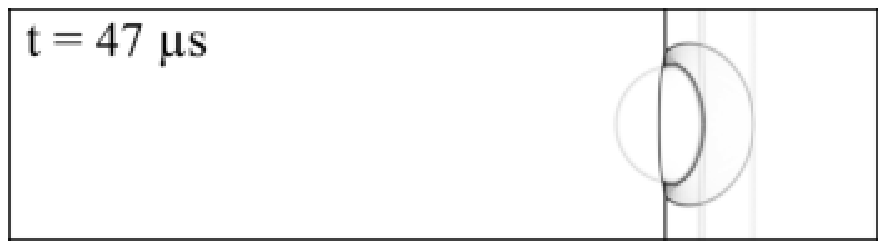}
\includegraphics[width=0.325\textwidth]{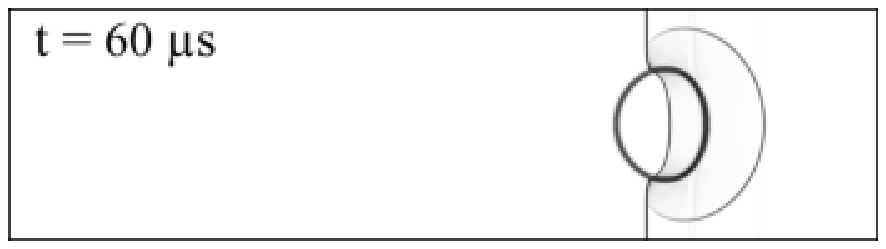}\\
\includegraphics[width=0.325\textwidth]{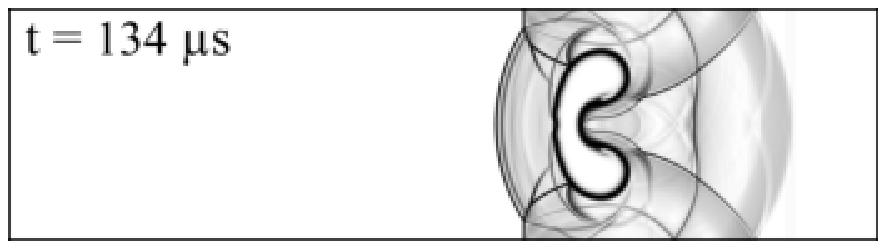}
\includegraphics[width=0.325\textwidth]{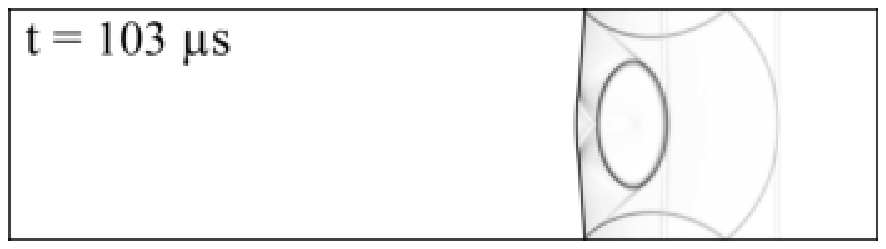}
\includegraphics[width=0.325\textwidth]{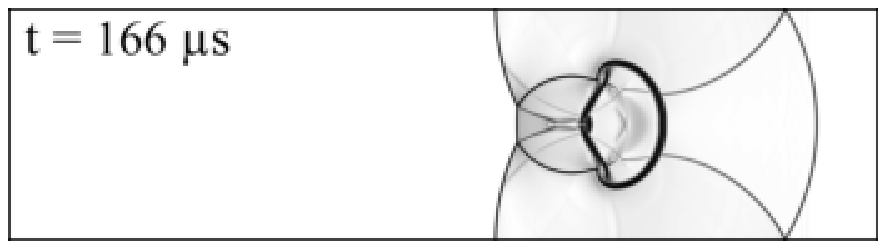}\\
\includegraphics[width=0.325\textwidth]{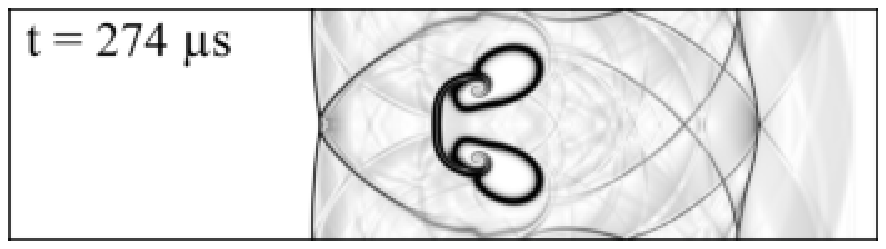}
\includegraphics[width=0.325\textwidth]{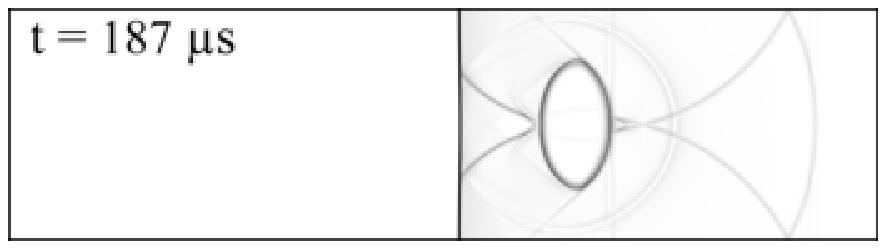}
\includegraphics[width=0.325\textwidth]{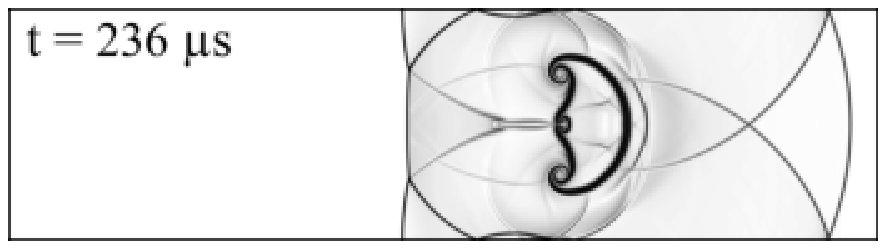}\\
\includegraphics[width=0.325\textwidth]{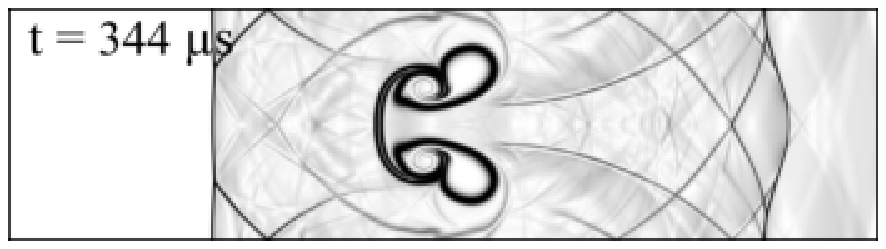}
\includegraphics[width=0.325\textwidth]{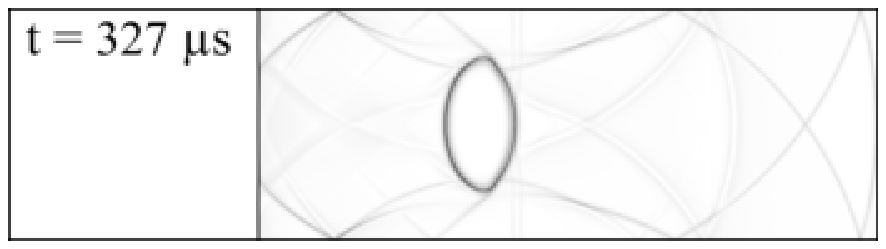}
\includegraphics[width=0.325\textwidth]{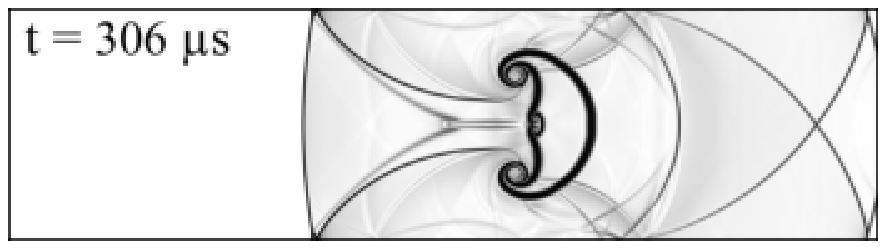}\\
\includegraphics[width=0.325\textwidth]{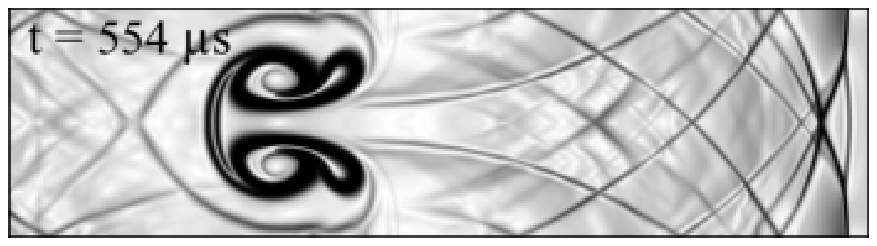}
\includegraphics[width=0.325\textwidth]{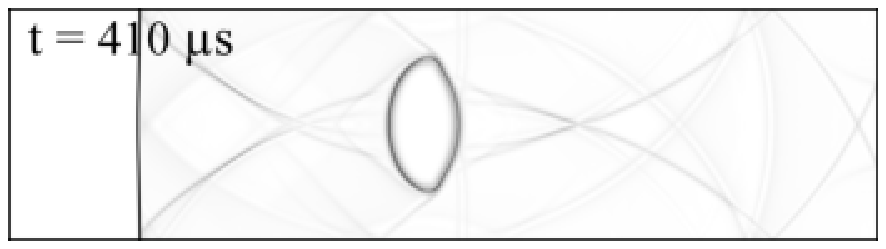}
\includegraphics[width=0.325\textwidth]{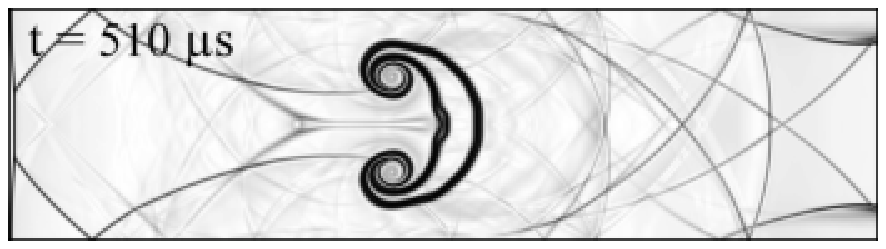}\\
\caption{Numerical schlieren images showing the interface 
evolution in time of the different air/bubble constitutions: 
air/He (left), air/N$_{2}$ (middle) and air/Kr (right) at $\textit{Ma} = 1.5$.}
\label{Fig8}
\end{figure*}

\begin{figure*}
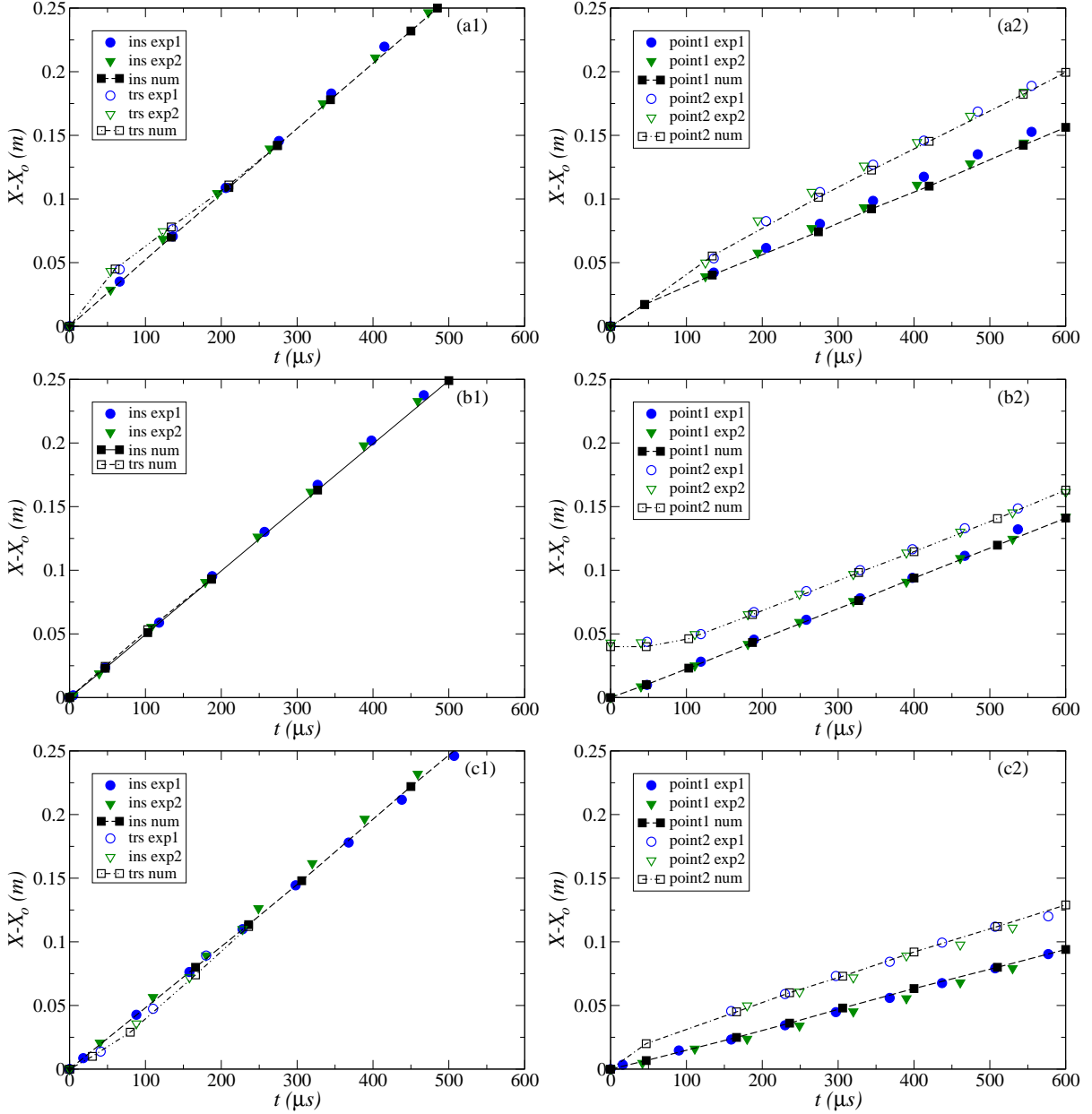

\includegraphics[width=0.48\textwidth]{Fig9a.eps}
\includegraphics[width=0.48\textwidth]{Fig9b.eps}\\
\includegraphics[width=0.48\textwidth]{Fig9c.eps}
\includegraphics[width=0.48\textwidth]{Fig9d.eps}\\
\includegraphics[width=0.48\textwidth]{Fig9e.eps}
\includegraphics[width=0.48\textwidth]{Fig9f.eps}\\
\caption{Incident shock (ins) and transmitted shock (trs) waves positions (left) 
and bubble  points 1 and 2 locations (right) at $\textit{Ma} = 1.5$ 
for air/He (a1, a2), air/N$_2$ (b1, b2) and air/Kr (c1, c2) constitutions. 
Comparison between present numerical results and experimental 
data~\cite{Layes2007}.}
\label{Fig9}
\end{figure*}

The positions of the characteristic interface points are recorded against time
in Fig.~\ref{Fig9}. This figure also shows the changing positions of the incident 
and transmitted shock waves for air/He, air/N$_{2}$ and air/Kr configurations along 
the tube, which are measured from the shock initial position $X_0$, Fig.~\ref{Fig5}.
The dynamic evolution of a bubble is observed by tracking points ($1$) and ($2$) 
originally placed on its contour, see Fig.~\ref{Fig5}. Point ($1$) is associated with 
the upstream front position and point ($2$) is the most downstream interface point. 
The usage of these tracking points to record numerically calculated spatial positions 
follows directly the experimental convention described in~\cite{Layes2007}. As it was  
earlier mention reproducible experiments with the same size of bubble were difficult 
to obtain. For example, the initial diameter of the nitrogen bubble in the experimental 
data is slightly larger than $4$~cm, Fig.~\ref{Fig9}(b2). Therefore the data from two 
realizations of the same experimental procedures presented in~\cite{Layes2007}, 
are utilized in the present comparison study. In spite of these difficulties 
the quantitative analysis of the computed positions in Fig.~\ref{Fig9} shows excellent 
agreement with the experimental findings. The numerical results confirm the validity 
of the underlying governing equations and numerical method. 

The first study presented in Fig.~\ref{Fig9}(a1, a2) shows the results for 
the helium bubble with lower acoustic impedance than the surrounding air. 
The difference in densities and therefore the higher sound speed 
in helium ($1007$~m/s) than in air ($340$~m/s) results in a higher speed for the 
transmitted shock through the helium bubble than for the incident shock in the air. 
The waves merge after passing the bubble to form a normal shock wave at around 
$270$~$\mu\rm{s}$ Fig.~\ref{Fig9}(a1). The early stages of the physical process reproduced 
numerically confirm the vorticity generation by the baroclinic effects. The rear interface 
of the helium bubble is caught by the front interface and the bubble evolves into a kidney 
shape. A penetrating high velocity jet along the flow direction moves through the bubble 
forming two symmetric flow configurations, Fig.~\ref{Fig8}. When the bubble deforms 
the associated flow field is subsequently split into two rings of vorticity. 
This characteristic separation further intensifies the deformation of the inhomogeneity.

The second study presented in Fig.~\ref{Fig9}(b1, b2) characterizes the interaction 
of a shock wave encountering a nitrogen bubble with comparable acoustic 
impedance as the surrounding air. The nitrogen and the air densities have 
similar values and therefore the corresponding Atwood number is close to zero. 
In such flow regime both the incident and the transmitted shock waves propagate 
with a small difference in velocities. After approximately $180$~$\mu\rm{s}$ 
the waves are combined again to form a planar shock wave. The generation and 
subsequent development of the vorticity field is negligible in this case. 
The compression process dominates the flow and the bubble evolution. The shape 
of the nitrogen bubble does not change significantly with time after around 
$200$~$\mu\rm{s}$ when the compression rate stabilizes.

The third study included in Fig.~\ref{Fig9}(c1, c2) reveals the numerical results 
for the krypton bubble. The krypton acoustic impedance is higher than the air
acoustic impedance. Such situation makes the transmitted shock through 
the krypton bubble moving more slowly than the incident 
shock in the surrounding air. These waves fully converge after $300$~$\mu\rm{s}$. 
This case clearly shows that the vorticity drives the distortion mechanism. 
The shock passage generates vorticity on the bubble interface owing to misalignment 
of the pressure and density gradients across the interface. 
The vortical flow then distorts the bubble interface together with 
a penetrating jet that is generated after around $160$~$\mu\rm{s}$ 
along the symmetry line of the bubble which moves upstream towards 
the right hand side. In all these cases the different times at which 
shocks leave the tube were recorded. The accelerated shock in the 
helium case left the tube after around 480~$\mu\rm{s}$, 
in the nitrogen case the shock left
the tube at around $500$~$\mu\rm{s}$ and finally, in the krypton case the shock was 
decelerated and left the tube after $510$~$\mu\rm{s}$.

%%%%%%%%%%%%%%%%%%
\subsubsection{Experiments of Zhai \it{et al.}~\cite{Zhai2011}}
%%%%%%%%%%%%%%%%%%
The comparison study with both experimental and numerical results was 
carried out for the interaction of a weak $\textit{Ma}=1.2$ 
planar shock wave with a sulphur hexafluoride SF$_6$ bubble immersed 
in air. The Atwood number for this configuration is $A=0.66$. 
The advantage of the considered experiment over the previous investigations 
of~\cite{Layes2007} is the application of a high speed schlieren photography 
with higher time resolution. This allows precise validations of the location 
of the wave front evolution both inside and outside the gas bubble at the very 
early stages of the interaction. 
As in~\cite{Layes2007} the experiment was performed using a rectangular shock 
tube, Fig.~\ref{Fig5}, but with different dimensions of the observation window 
which were $0.07\times0.5$~$\rm{m}$. The bubble initial diameter 
is $D_0=0.03~\rm{m}$ and the centre is located at a $0.02~\rm{m}$
distance from the shock. The numerical initial conditions are set per analogy 
to the previous investigation and are summarised in Table~\ref{Table5}. 
The $2000\times560$ computational grid provided a resolution 
of $240$ cells per bubble diameter. The CFL number was set to 
be equal to $0.3$.

\begin{table}
\caption{Initial conditions of the air/SF$_6$ shock-bubble interaction test}
\center
\begin{ruledtabular}
\begin{tabular}{lccc}
Physical property & SF$_6$ bubble &pre-shocked air &post-shocked air\\
\hline
Density, $kg/m^{3}$  &  $5.97$  &  $1.19$ &  $1.597$\\
Horizontal velocity, $m/s$  & $0.0$  &  $0.0$ &  $105.6$\\
Sound speed, $m/s$  & $134$  &  $346$ & $367$\\
Pressure, $Pa$ & $101325$ &  $101325$ &  $153339$\\
Heat capacity ratio, $\gamma$ &  $1.1$  &  $1.4$ &  $1.4$\\
Acoustic impedance, $Pa\cdot s/m$ & 799.98 & 411.74 & 586.09
\end{tabular}
\end{ruledtabular}
\label{Table5}
\end{table} 	

\begin{figure}
\centering
\includegraphics[width=0.19\textwidth]{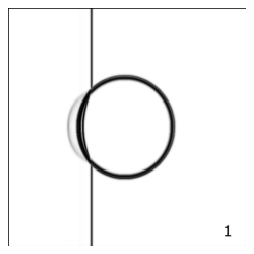}
\includegraphics[width=0.19\textwidth]{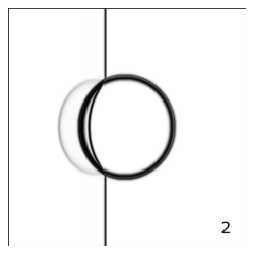}
\includegraphics[width=0.19\textwidth]{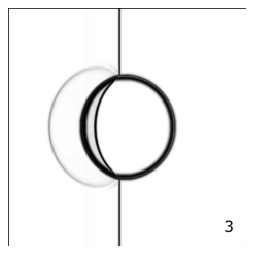}
\includegraphics[width=0.19\textwidth]{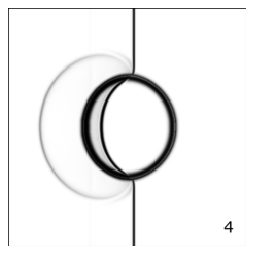}
\includegraphics[width=0.19\textwidth]{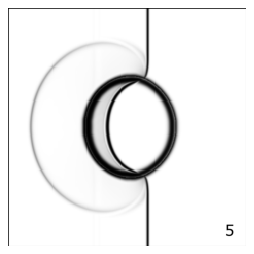}\\
\includegraphics[width=0.19\textwidth]{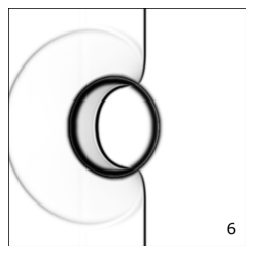}
\includegraphics[width=0.19\textwidth]{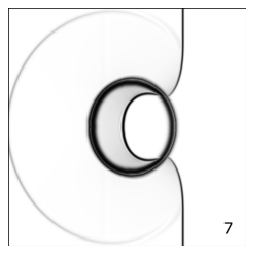}
\includegraphics[width=0.19\textwidth]{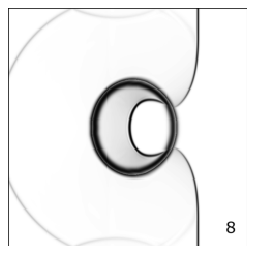}
\includegraphics[width=0.19\textwidth]{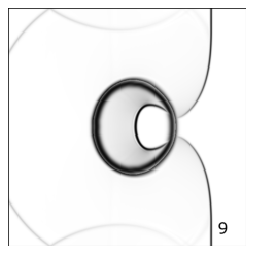}
\includegraphics[width=0.19\textwidth]{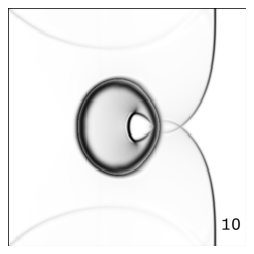}\\
\includegraphics[width=0.19\textwidth]{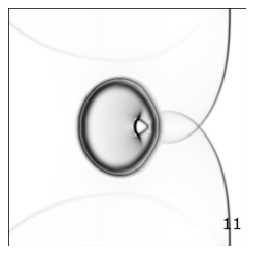}
\includegraphics[width=0.19\textwidth]{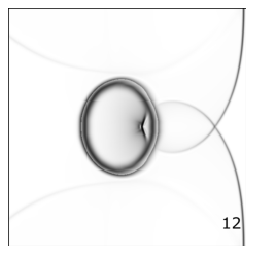}
\includegraphics[width=0.19\textwidth]{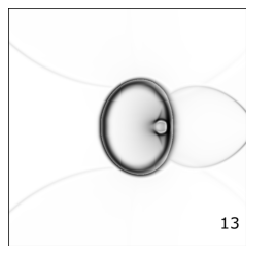}
\includegraphics[width=0.19\textwidth]{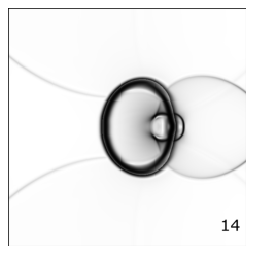}
\includegraphics[width=0.19\textwidth]{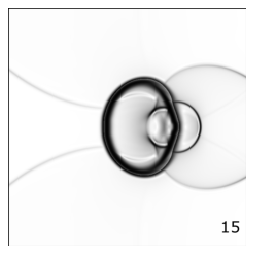}\\
\caption{Sequence of schlieren images representing the SF$_6$ 
bubble evolution as a result of its interaction with a planar
shock wave. The time interval between successive images
from the numerical simulation is $10$~$\mu\rm{s}$.}
\label{Fig10}
\end{figure}

The schlieren images from the present simulation 
are collected in Fig.~\ref{Fig10}. 
The time interval between consecutive images is set to 
$10$~$\mu\rm{s}$ to capture the same sequences of the process 
as presented earlier in the experiment of~\cite{Zhai2011}. The 
images reveal the characteristic moments of the interaction and 
are in a very good agreement with the experimental and
numerical results of~\cite{Zhai2011}. The images are numbered 
using the same convention as in the reference paper.
The transmitted shock wave takes the convergent shape owing to 
the difference in acoustic impedance, Fig.~\ref{Fig10} (images 1 to 5). 
Images 6 to 9, in the same figure, show two parts of the incident shock 
wave passing the top and the bottom poles of the bubble and moving towards 
the most downstream point of the interface. The transmitted shock starts 
to converge inside the bubble towards the centre of the downstream interface, 
Fig.~\ref{Fig10} (images 10 to 13). As a result the formation of 
the penetrating jet can be observed in the following images. The process 
is driven by a high pressure zone resulting from the shock formation which 
concentrated at the downstream pole, Fig.~\ref{Fig10} (images 14 and 15). 
This causes an explosion producing a refracted shock wave moving through 
the downstream boundary of the bubble from left to right and a shock wave 
propagating inside the bubble.

\begin{figure}
\centering
\begin{minipage}[t]{0.45\linewidth}
\includegraphics[width=0.75\textwidth]{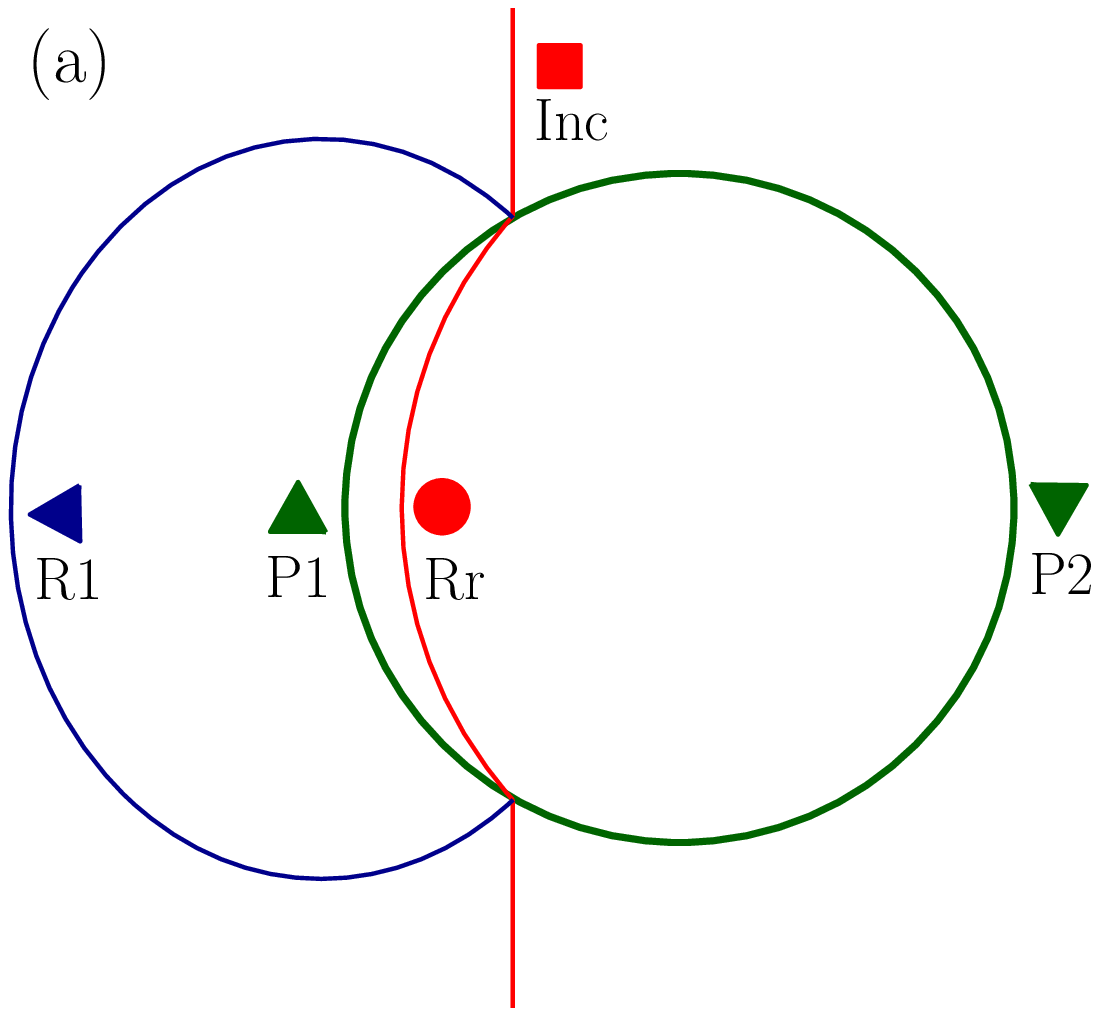}
\end{minipage}
%\qquad
%\hfill
\begin{minipage}[t]{0.45\linewidth}
\raisebox{1cm}{\includegraphics[width=0.75\textwidth]{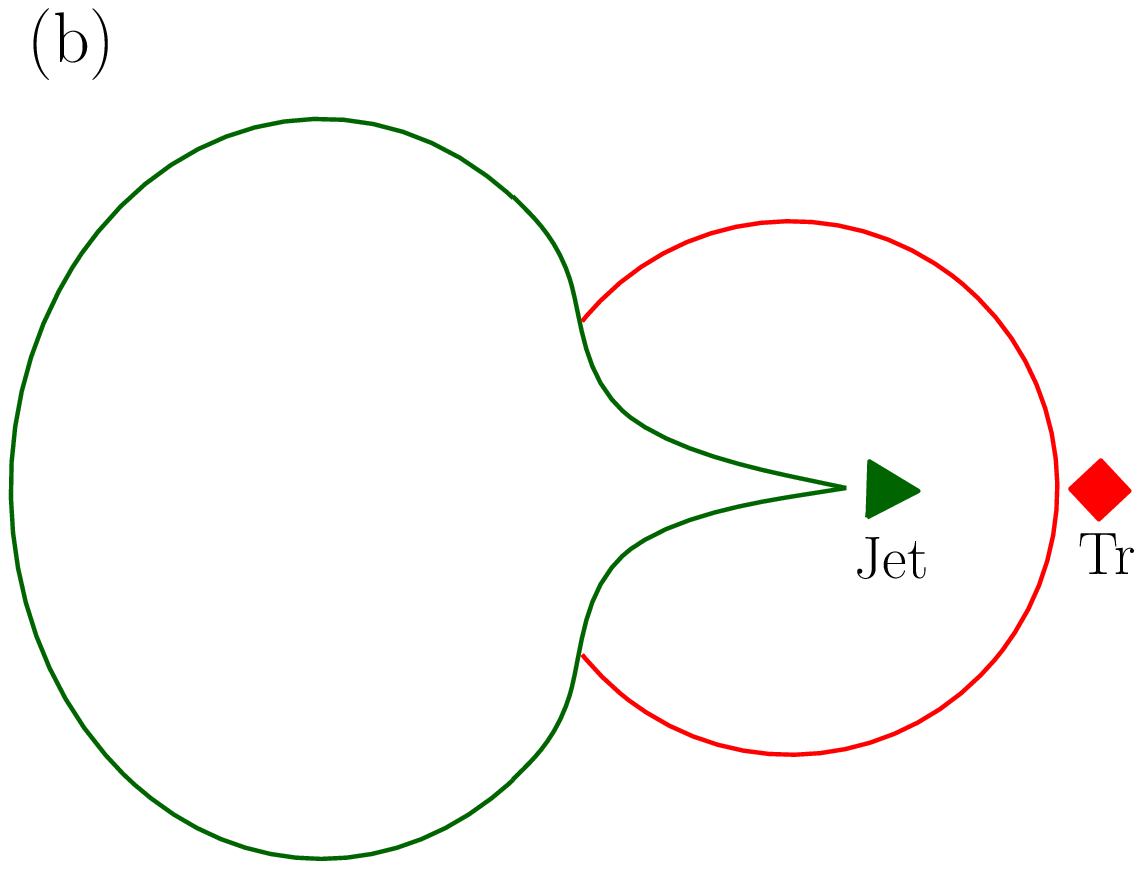}}
\end{minipage}
\caption{Schematic diagram with the characteristic elements of the interaction 
of a shock with a SF$_6$ bubble. Evolution stages: (a) early 
stage of the interaction (b) a moment after the shock wave passed 
the bubble. Tracking points: Inc-incident shock, 
P1-upstream interface, P2-downstream interface, 
R1-reflected shock, Rr-refracted shock, Tr-transmitted shock and Jet head}
\label{Fig11}
\end{figure}

The evolution is recorded using $x-t$ diagrams. Figure~\ref{Fig11} 
defines the different tracking locations for the considered 
shock-bubble interaction reproduced in Fig.~\ref{Fig10}.
The positions of the upstream interface~(P1),
downstream interface~(P2), refracted~(Rr), reflected~(R1) and transmitted 
shock~(Tr) are obtained at the horizontal axis while the incident shock~(Ins) 
wave is measured at the undisturbed locations above the bubble.

\begin{figure}
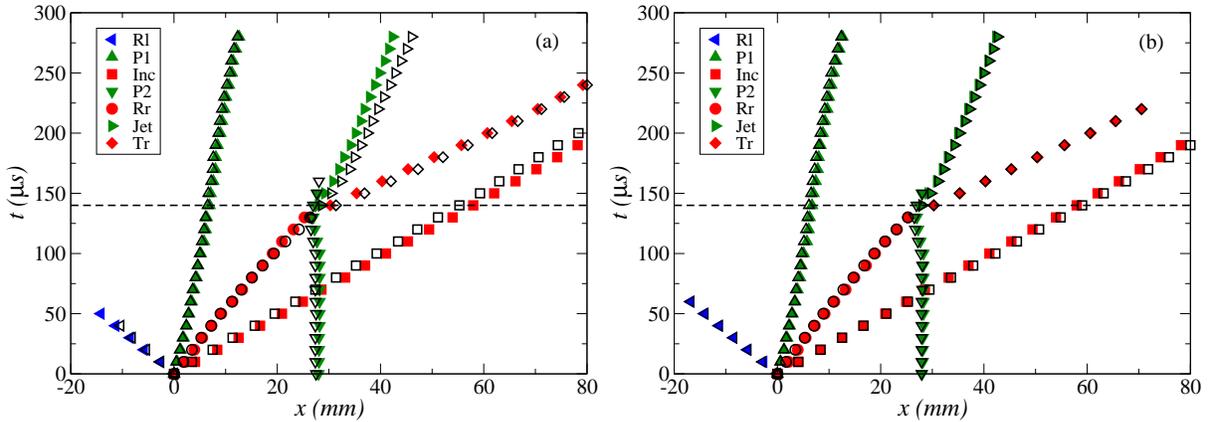

\centering
\includegraphics[width=0.48\textwidth]{Fig12a.eps}
\includegraphics[width=0.48\textwidth]{Fig12b.eps}\\
\caption{Recorded various shocks and interface positions during 
the interaction of a shock wave with a SF$_6$ bubble. Symbols are linked
to the tracking locations shown in Fig.~\ref{Fig11}. The present numerical
results (filled symbols) are compared with the experimental (a), 
and the numerical predictions (b) of~\cite{Zhai2011}.}
\label{Fig12}
\end{figure}

Figure~\ref{Fig12} refers to the distinct tracking points indicated
in Fig.~\ref{Fig11}. The positions of these characteristic points, 
representing interface and various waves involved, are determined 
during the numerical simulation and their evolution is compared 
with (a) the experimental and (b) the numerical results 
reported in~\cite{Zhai2011}. The numerical predictions are in perfect 
agreement with the experimental data in the first stages of the interaction
although a small difference in the position of the most downstream pole of 
the bubble is observed.   

Table~\ref{Table6} lists the velocities associated with the different shock waves. 
$V_{jet}$ is the velocity of the jet and $t_{jet}$ refers to the time at which 
the jet starts to form. Slight differences can be noticed between the numerical 
and experimental results. The main reason for these differences is due to the 
impurity of the gases inside and outside the bubble in the experiment which was 
acknowledged by~\cite{Zhai2011}.

\begin{table}
\caption{Comparison of the different velocities in the air/SF$_6$ 
shock bubble interaction test.}
\center
\begin{ruledtabular}
\begin{tabular}{lccccccc}
    &  {\it Ma}   &  $V_{inc}$ &  $V_{Rr}$ &  $V_{Rl}$ &  $V_{Tr}$ &  $V_{jet}$ &  $t_{jet}$\\
\hline
Experiment~\cite{Zhai2011} & $1.2$ & $395$ & $203$ & $347$ & $488$ & $114$ & $140$\\
Computation~\cite{Zhai2011} & $1.2$ & $415$ & $194$ & $380$ & $501$ & $108$ &  $140$\\
Current computation       & $1.2$  & $410$ &  $197$ &  $380$ &  $501$ & $108$ & $140$\\
\end{tabular}
\end{ruledtabular}
\label{Table6}
\end{table}

%%%%%%%%%%%%%%%%%%%%%%%%%%%%%%%%%%%%%%%%%%%%%%%%%%%%%%%%%
\subsection{Interface evolution and vorticity production 
as a function of Mach and Atwood numbers} 
\label{Mach_Atwood_comparison}
%%%%%%%%%%%%%%%%%%%%%%%%%%%%%%%%%%%%%%%%%%%%%%%%%%%%%%%%%
 
After these successful validations the numerical procedures are applied to
examine additional cases for which experimental data 
cannot be collected owing to the restrictions set by the physical apparatus. 
These new computational simulations consider the effect of a wider range 
of Atwood numbers on the shape of the interface as well as the effect 
of the Mach number on the interface growth and development. The influence
of the Atwood and Mach number changes on the baroclinic source
of the vorticity field is also investigated. Apart from the gases considered 
in the previous section the extra cases include the pairings of air/argon (Ar) 
with $A = 0.13$. Therefore the numerical study accounts for the total of five
different bubble/air configurations interacting with the waves for which Mach 
numbers were set to be 1.5,~2,~2.5 and 3. The initial data for these 
simulations are listed in Tables~\ref{Table7} and~\ref{Table8}. 

\begin{table}
\caption{\label{Table7}Properties of the gas bubbles at atmospheric 
pressure and 15$^o\rm{C}$.}
\begin{ruledtabular}
\begin{tabular}{lccccc}
Physical property   & He & N$_{2}$ & Ar & Kr & SF$_{6}$\\
\hline
Density, $kg/m^3$   &0.169 &1.19 & 1.67  & 3.55 & 6.27 \\
Sound speed, $m/s$  &1000 &345 &318 &218 & 132 \\
Heat capacity ratio $\gamma$ & 1.664 & 1.40 & 1.664  & 1.67 & 1.08\\
Acoustic impedance, $Pa\cdot s/m$ & 169 & 411 & 531 & 775 & 828
\end{tabular}
\end{ruledtabular}
\end{table}

\begin{table*}
\caption{\label{Table8}Properties of air in the high pressure chamber.}
\begin{ruledtabular}
\begin{tabular}{lllll}
Property  & Case I & Case II & Case III & Case IV \\
\hline
Density, $kg/m^3$ & 2.242 & 3.211 & 4.013 & 4.644\\
Pressure, $Pa$ & $2.49091\times10^5$ &$4.55963\times10^5$ &$7.21941\times10^5$ &$1.047025\times10^6$\\
Mach number  &1.5 &2 & 2.5 & 3 \\
\end{tabular}
\end{ruledtabular}
\end{table*} 

\begin{figure*}
\includegraphics[width=0.19\textwidth]{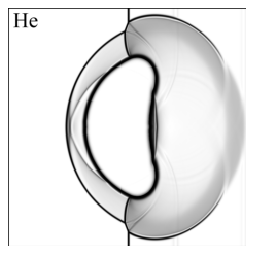}
\includegraphics[width=0.19\textwidth]{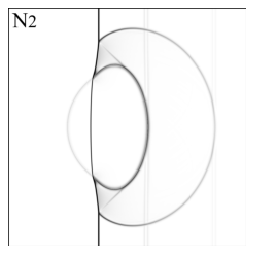}
\includegraphics[width=0.19\textwidth]{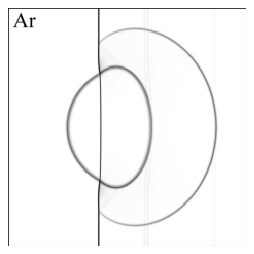}
\includegraphics[width=0.19\textwidth]{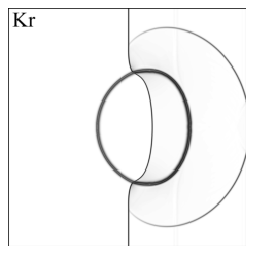}
\includegraphics[width=0.19\textwidth]{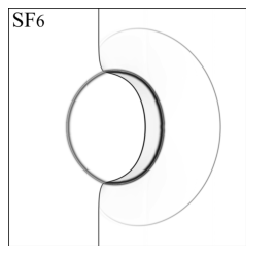}
\caption{Numerical schlieren images for the various air/gas constitutions
at time $60$~$\mu\rm{s}$ and $\textit{Ma} = 1.5$.}
\label{Fig13}
\end{figure*}
\begin{figure}
\includegraphics[width=0.5\textwidth]{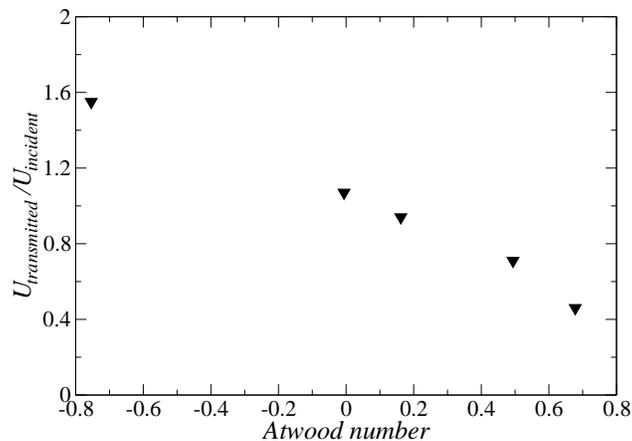}
\caption{Variation of the normalised transmitted shock velocity as a function
of the Atwood number at time $60$~$\mu\rm{s}$ and $\textit{Ma} = 1.5$.}
\label{Fig14}
\end{figure}
\begin{figure*}
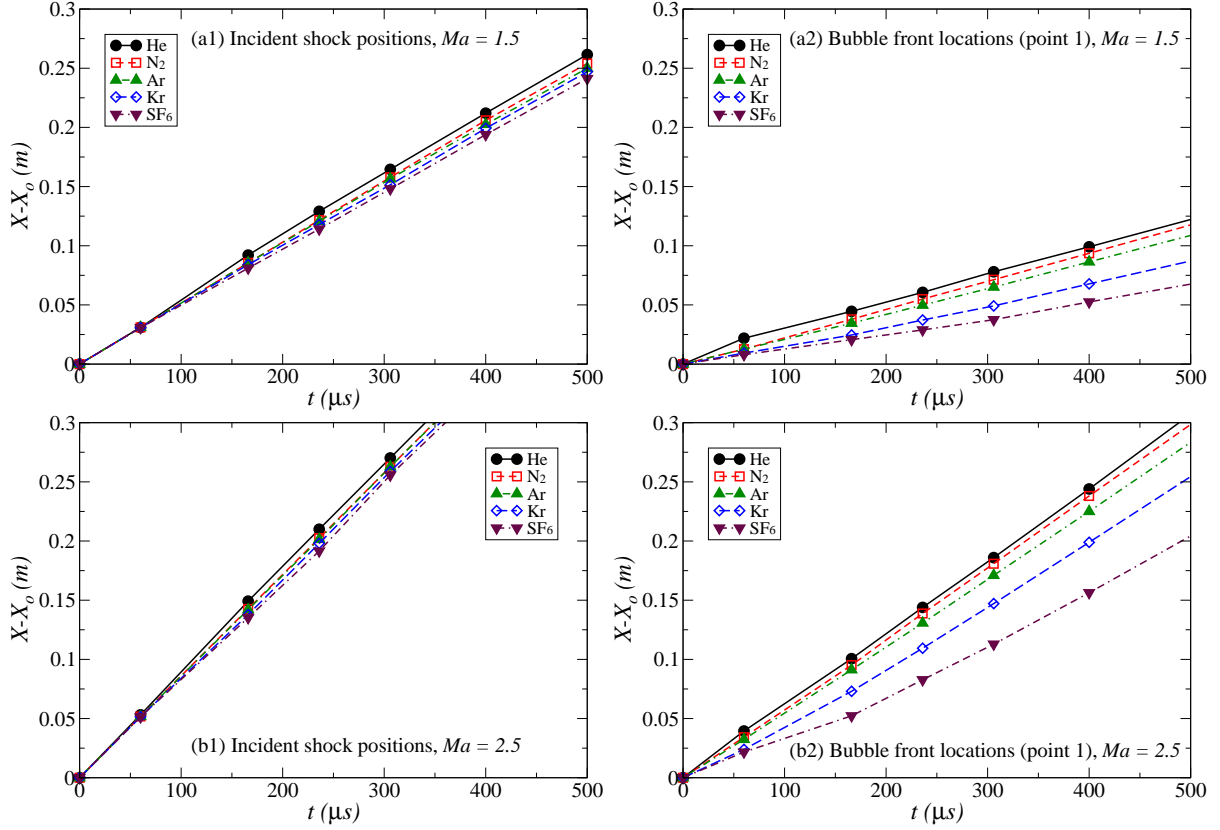

\includegraphics[width=0.48\textwidth]{Fig15a.eps}
\includegraphics[width=0.48\textwidth]{Fig15b.eps}\\
\includegraphics[width=0.48\textwidth]{Fig15c.eps}
\includegraphics[width=0.48\textwidth]{Fig15d.eps}\\
\caption{Incident shock and bubble upstream point 1 positions for different
air/gas constitutions at $\textit{Ma} = 1.5$ (a1, a2)
and $\textit{Ma} = 2.5 $ (b1, b2).}
\label{Fig15}
\end{figure*}
 
\begin{figure*}
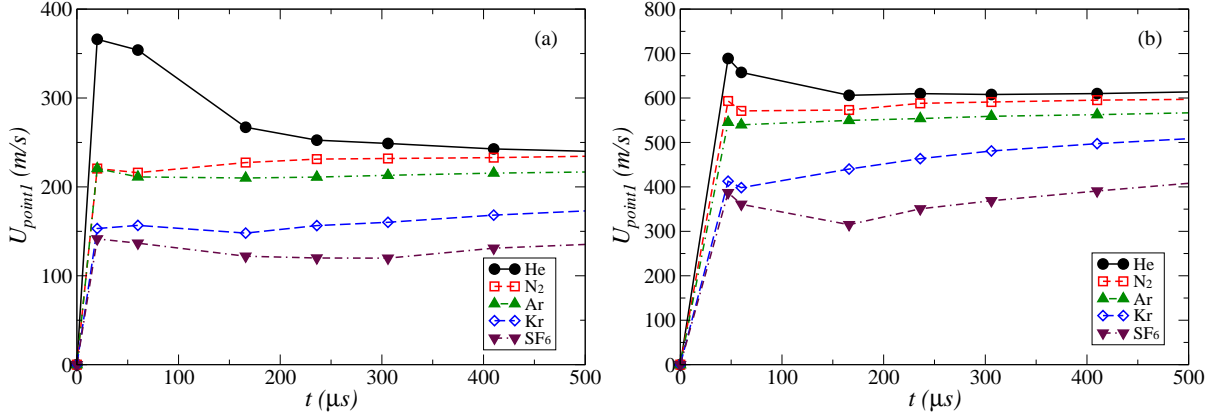

\includegraphics[width=0.48\textwidth]{Fig16a.eps}
\includegraphics[width=0.48\textwidth]{Fig16b.eps}\\
\caption{The velocities of point (1) for different gas bubbles 
as a function of time for (a) ${\it Ma} = 1.5$ and (b) ${\it Ma} = 2.5$.}
\label{Fig16}
\end{figure*}

Figure~\ref{Fig13} shows the mixture density field profiles for different air/gas
constitutions captured at the same early stage (60~$\mu\rm{s}$) of the bubble
interactions with a shock wave of $\textit{Ma} = 1.5$. The form of deformation
of the bubble during the penetration of the shock wave is determined by the 
density and acoustic impedance of each constituent. The fastest speed of
penetration was observed in the He bubble and this speed was at the same 
time faster than the normal incident shock. The situation was different
in the cases of N$_{2}$ and Ar, where a slight difference between the speeds
of transmitted and incident shock waves results in a smaller deformation
of the bubble. The case of Kr and SF$_{6}$ exemplified a scenario opposite
from that for He. In these cases the transmitted shock propagates through
the bubbles more slowly than the incident shock outside the bubbles boundary.
The transmitted shock in the SF$_{6}$ bubble moves also more slowly than 
in the Kr bubble. The early stages of the shock-bubble interaction in the 
last two cases did not allow for large deformations of the bubble.
Figure~\ref{Fig14} illustrates the relation between the velocity ratio 
($U_{transmitted}/U_{incident}$) and the Atwood number at time 
$60$~$\mu\rm{s}$. It is observable that as the Atwood number goes towards
positive values and becomes larger, the transmitted shock propagates 
through the bubble more slowly. Figure~\ref{Fig15} shows the incident shock 
position and the location of the bubbles, filled with different gases, along
the domain as a function of time for two different Mach numbers: 
$\textit{Ma} = 1.5$ and $\textit{Ma} = 2.5$. It is confirmed in 
Fig.~\ref{Fig15}(a1) and (b1) that the incident shock travels through
the domain containing light bubbles faster than in the cases of heavy bubbles. 
The fact that the transmitted shock could accelerate or decelerate 
in the bubble environment has consequences at a later time, 
when the transmitted shock leaves the bubble and eventually combines with 
the incident shock. These are manifested by higher or lower wave speeds as
compared to the medium containing a bubble with comparable physical properties
to the surrounding medium or not containing a bubble at all. The location of
the bubbles has been measured by tracking point ($1$) on the front of the 
bubble (upstream side, Fig.~\ref{Fig5}). The interpretation of 
Figs.~\ref{Fig15}(a2) and (b2) confirms that as the bubble is heavier
it moves more slowly. Similarly these figures show
the effect of the Mach number on the movement of the gas bubbles. 
The bubble covers a longer distance with higher Mach numbers.
To assist in understanding the changes in the dynamics of the interface, 
the velocity values associated with point 1 on the upstream pole of the 
gas bubble were monitored. The values were collected for the Mach numbers 
$1.5$ and $2.5$, Fig.~\ref{Fig16}.

\begin{figure*}
\includegraphics[width=0.244\textwidth]{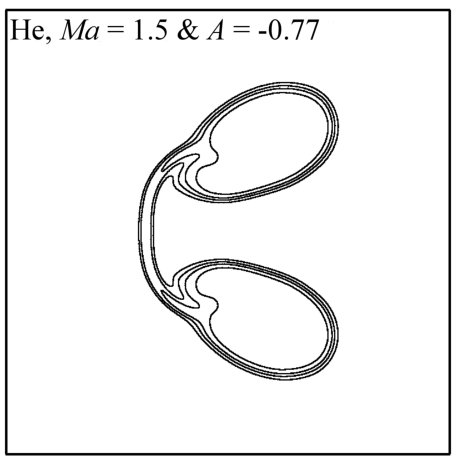}
\includegraphics[width=0.244\textwidth]{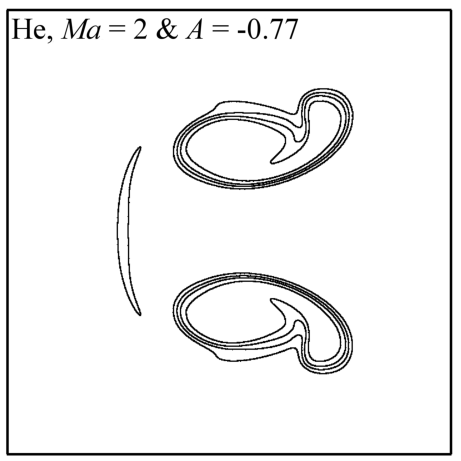}
\includegraphics[width=0.244\textwidth]{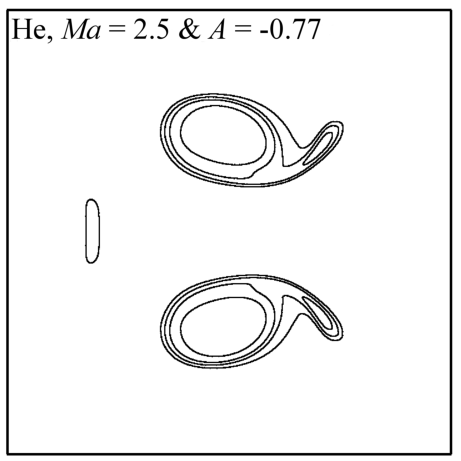}
\includegraphics[width=0.244\textwidth]{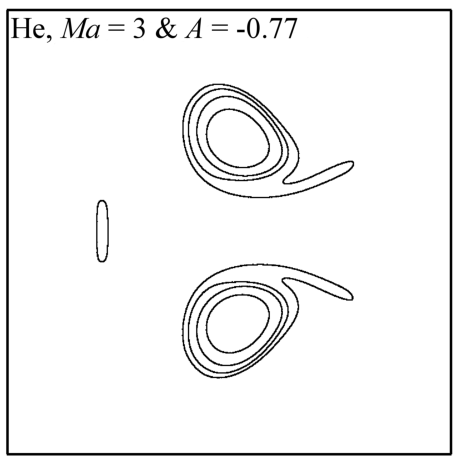}\\
\includegraphics[width=0.244\textwidth]{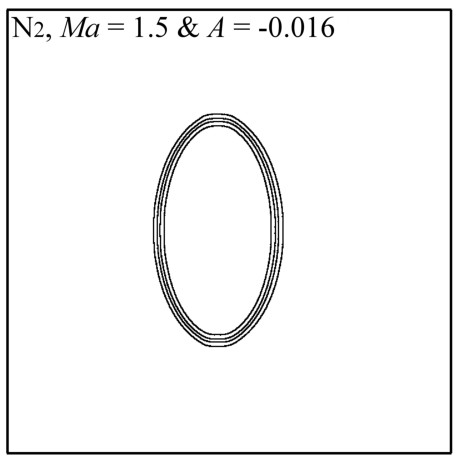}
\includegraphics[width=0.244\textwidth]{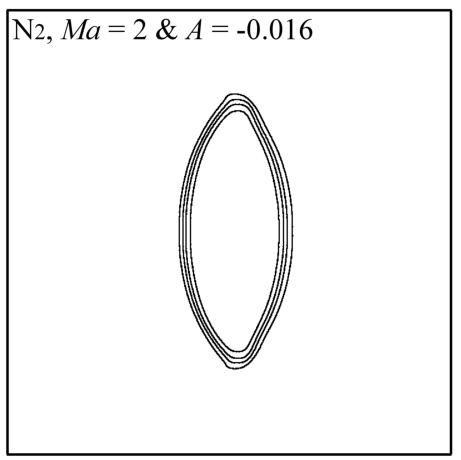}
\includegraphics[width=0.244\textwidth]{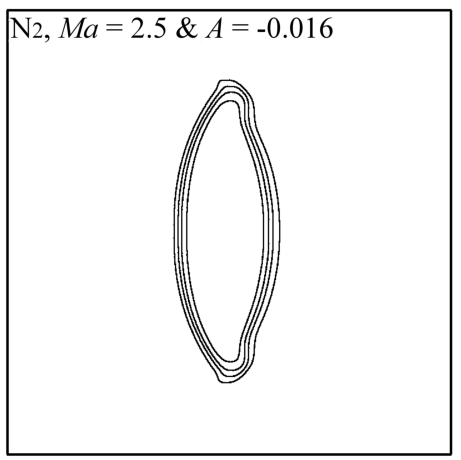}
\includegraphics[width=0.244\textwidth]{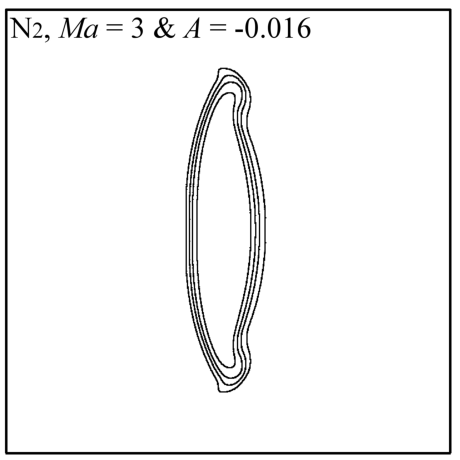}\\
\includegraphics[width=0.244\textwidth]{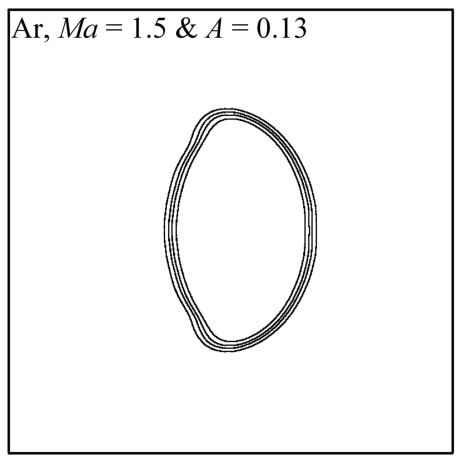}
\includegraphics[width=0.244\textwidth]{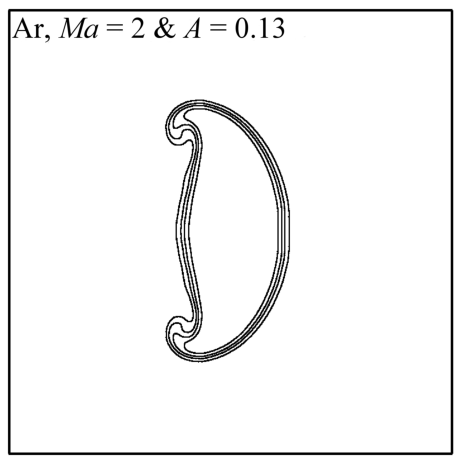}
\includegraphics[width=0.244\textwidth]{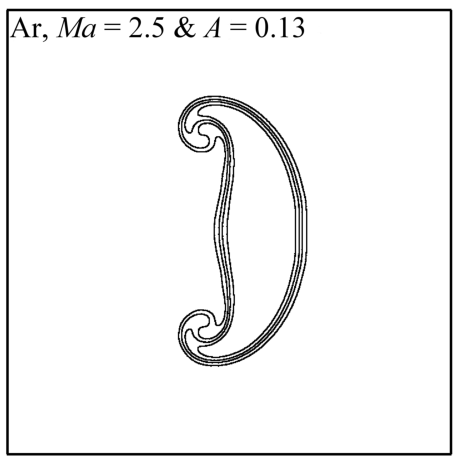}
\includegraphics[width=0.244\textwidth]{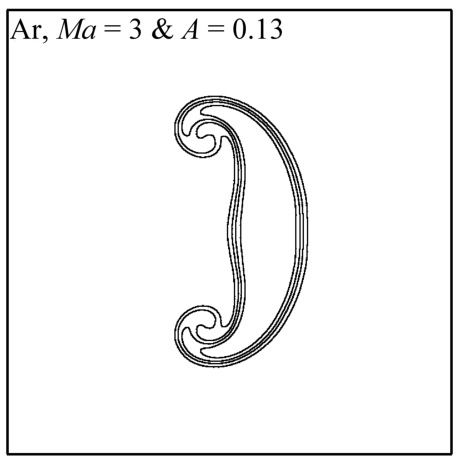}\\
\includegraphics[width=0.244\textwidth]{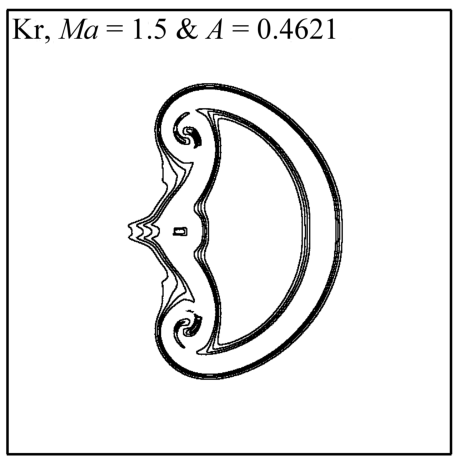}
\includegraphics[width=0.244\textwidth]{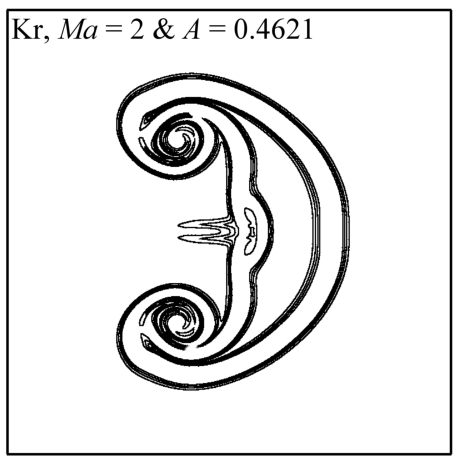}
\includegraphics[width=0.244\textwidth]{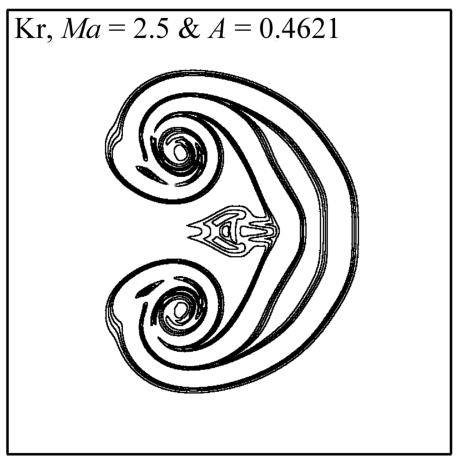}
\includegraphics[width=0.244\textwidth]{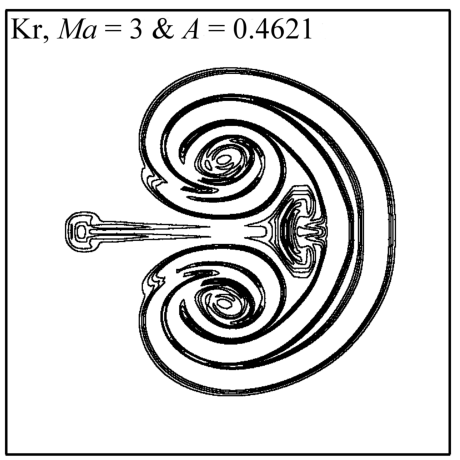}\\
\includegraphics[width=0.244\textwidth]{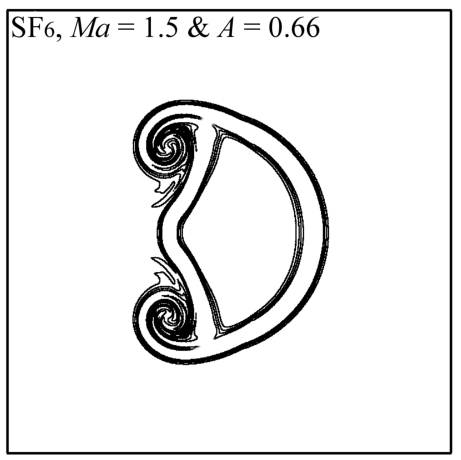}
\includegraphics[width=0.244\textwidth]{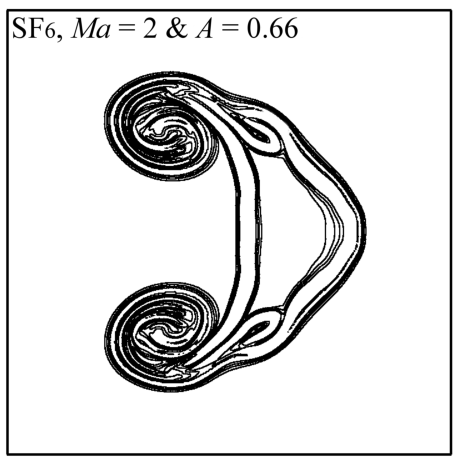}
\includegraphics[width=0.244\textwidth]{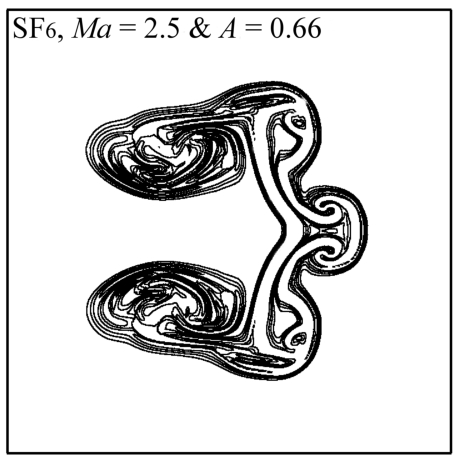}
\includegraphics[width=0.244\textwidth]{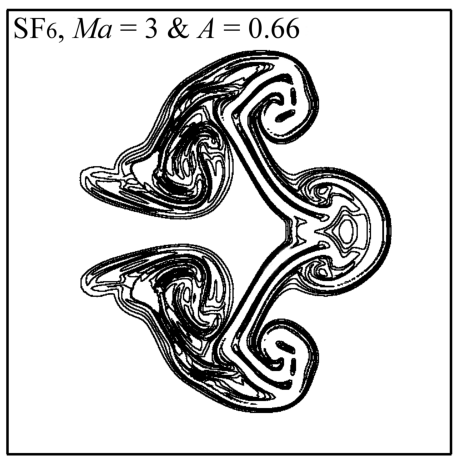}\\
\caption{Volume fraction contours of various gas/air 
constitutions and {\it Ma} numbers at time 236~$\mu\rm{s}$.}
\label{Fig17}
\end{figure*}

\begin{figure}
\includegraphics[width=0.48\textwidth]{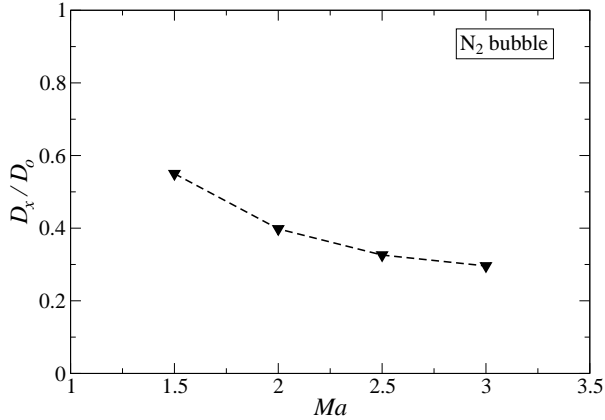}
\caption{Compression ratio of N$_2$ bubble as a function of the Mach 
number at time t = $510$~$\mu\rm{s}$.}
\label{Fig18}
\end{figure}

\begin{figure*}
\includegraphics[width=0.24\textwidth]{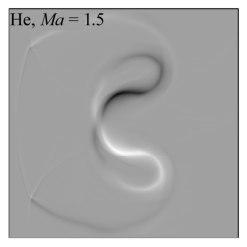}
\includegraphics[width=0.24\textwidth]{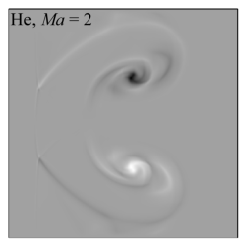}
\includegraphics[width=0.24\textwidth]{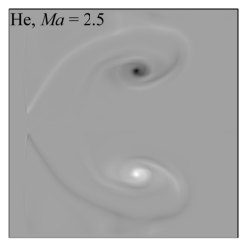}
\includegraphics[width=0.24\textwidth]{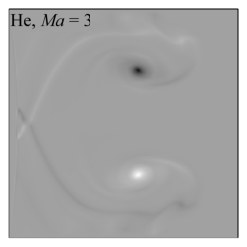}\\
\includegraphics[width=0.24\textwidth]{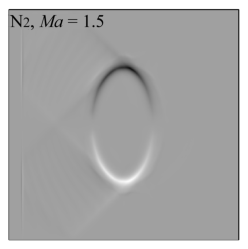}
\includegraphics[width=0.24\textwidth]{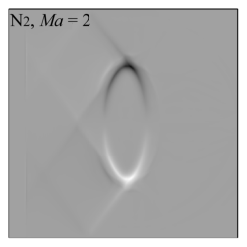}
\includegraphics[width=0.24\textwidth]{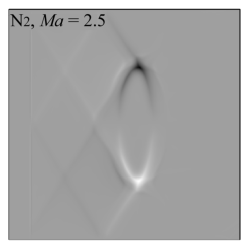}
\includegraphics[width=0.24\textwidth]{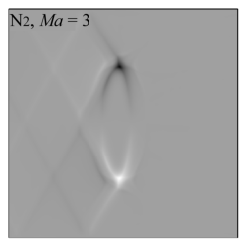}\\
\includegraphics[width=0.24\textwidth]{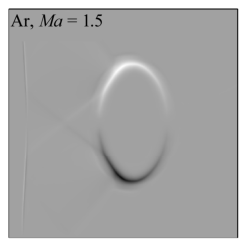}
\includegraphics[width=0.24\textwidth]{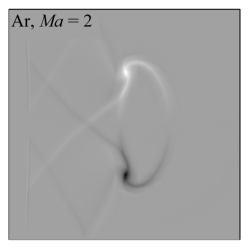}
\includegraphics[width=0.24\textwidth]{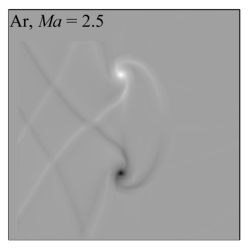}
\includegraphics[width=0.24\textwidth]{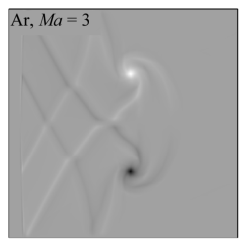}\\
\includegraphics[width=0.24\textwidth]{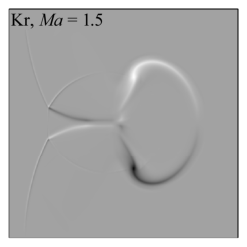}
\includegraphics[width=0.24\textwidth]{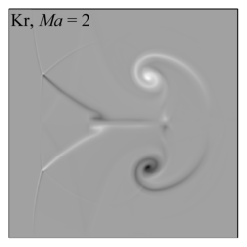}
\includegraphics[width=0.24\textwidth]{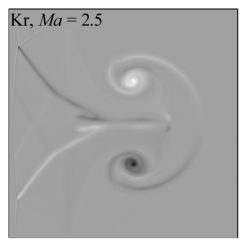}
\includegraphics[width=0.24\textwidth]{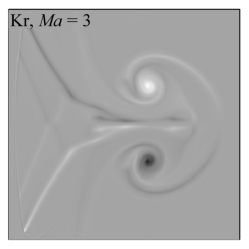}\\
\includegraphics[width=0.24\textwidth]{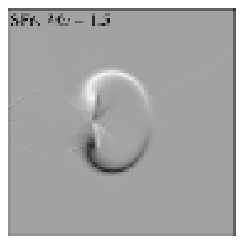}
\includegraphics[width=0.24\textwidth]{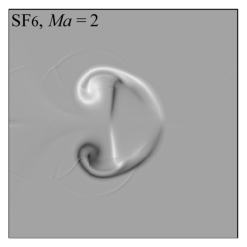}
\includegraphics[width=0.24\textwidth]{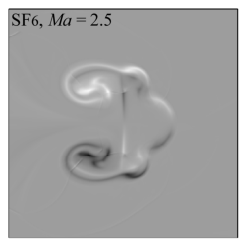}
\includegraphics[width=0.24\textwidth]{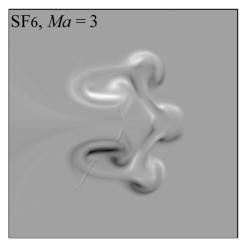}\\
\caption{Vorticity field contours of various gas/air 
constitutions and {\it Ma} 
numbers at time $166~\mu\rm{s}$.}
\label{Fig19}
\end{figure*}

Figure~\ref{Fig17} presents the evolutionary patterns of the interfaces 
represented by the volume fraction contours for various air/gas 
configurations and Mach numbers. The images are all taken at the same 
physical time equal to $236$~$\mu\rm{s}$ after the shock started to interact
with the gas bubbles. Figure~\ref{Fig17} can be read either from 
left to right (horizontal images - increase of Mach number) 
or from top to bottom (vertical images - increase of Atwood number). 
In the horizontal view one can see the effect of the Mach number on the 
interface evolution. The higher the Mach number the more changes in the 
interface shape, which is clearly seen in the last row of this horizontal 
view where the bubble undergoes distortion and consequently is divided 
into three entities with a significant interface evolution. In the case 
of the He bubble two symmetric contours can be observed as a result of 
the shock-bubble interaction. At the same time a high 
speed penetrating jet develops along the axis of symmetry in the main 
flow direction. The formation of the symmetric contours is more pronounced 
with the higher Mach numbers, leading eventually to a faster splitting of
the bubble into two entities. A different situation is observed in case of
the N$_2$ bubble. Here, the bubble is experiencing a compression process which
intensifies with the higher Mach number. It is also found that the compression 
process happens at the early stages of the shock-bubble interactions allowing
the bubble to stabilize its shape after around $200$~$\mu\rm{s}$ from 
the start of the interaction. This physical behaviour can be attributed 
to the fact that there is no penetrating jet or associated vorticity field 
(as the vorticity values are negligible in this case) which is clearly a direct
consequence of the small density ratio of the constituents. Figure~\ref{Fig18} 
illustrates the rate of compression of the N$_{2}$ bubble as a function 
of Mach number at the time $510$~$\mu\rm{s}$. The compression ratio increases
with the Mach number and it is measured by dividing the horizontal diameter 
of the bubble ($D_x$) at time $510$~$\mu\rm{s}$ by the initial diameter 
($D_o$). The Ar, Kr and SF$_{6}$ bubbles undergo a similar physical process 
until the moment when the baroclinic source of vorticity comes into play. 
This leads to greater bubble deformation and its interface distortion 
is even more apparent with the increasing Mach number. Another distinct 
feature of this process is the formation of the penetrating high speed jet 
along the bubble axis of symmetry, which moves in the opposite direction
to the normal shock wave. The interface changes and jet development are
clearer for a higher Atwood number Kr and SF$_{6}$ bubbles. The cases with 
the higher absolute value of the Atwood number experience a higher rate of
the bubble deformation with increasing Mach numbers. In contrast, 
for the Atwood number close to zero the deformation rate of the interface 
is relatively slow. 

\begin{figure}
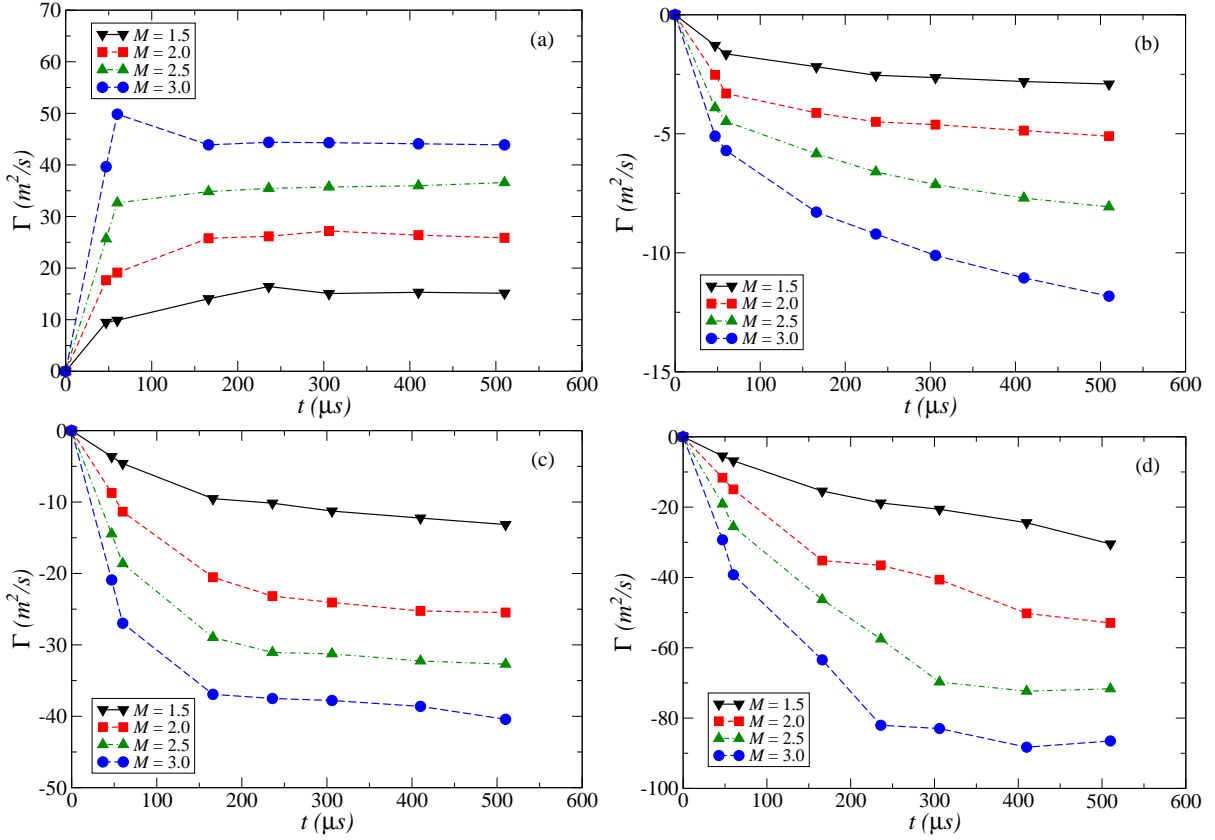

\includegraphics[width=0.48\textwidth]{Fig20a.eps}
\includegraphics[width=0.48\textwidth]{Fig20b.eps}\\
\includegraphics[width=0.48\textwidth]{Fig20c.eps}
\includegraphics[width=0.48\textwidth]{Fig20d.eps}
\caption{Time evolution of the circulation during the shock-bubble
interaction process for different Mach numbers and gases: 
(a) air/He,
(b) air/Ar, 
(c) air/Kr
and (d) air/SF$_{6}$
constitutions.}
\label{Fig20}
\end{figure}

\begin{table}
\caption{\label{Table9}Circulation {$\Gamma$(m$^2$/s)} 
at time t = $166$~$\mu\rm{s}$ for various flow constitutions 
and {\it Ma} numbers.}
\begin{ruledtabular}
\begin{tabular}{lrrrr}
$Ma$ & 1.5 & 2.0 & 2.5 & 3.0\\
\hline
air/He   &  7.583 &   8.869 &    9.537 &    9.755\\
air/N$_2$ & 0.184 &   0.096 &    0.097 &    0.113\\
air/Ar & $-1.820$ & $-2.453$ & $-2.621$ & $-2.828$\\
air/Kr & $-5.709$ & $-8.267$ & $-8.986$ & $-9.137$\\
air/SF$_6$ & $-11.649$ & $-17.413$ & $-23.209$ & $-26.113$
\end{tabular}
\end{ruledtabular}
\end{table}

Figure~\ref{Fig19} shows the development of the vorticity field
for all considered bubbles as a function of the Mach number. 
The snapshots were taken at the same time t = $166$~$\mu\rm{s}$. 
These pictures assist in the interpretation of the interface 
evolution process discussed previously, in which the vorticity 
creation plays an important role. To better understand Fig.~\ref{Fig19}, 
the vorticity field for different gases and Mach numbers is quantified 
by calculating the total circulation generated in the symmetrical half 
of the computational domain. The circulation values are listed
in Table~\ref{Table9} for  different Atwood and Mach numbers. 
The time evolutions of these values are presented in in Fig.~\ref{Fig20}. 
The shock propagates from the right to the left. Therefore in the 
case of light bubbles positive (anticlockwise) vortices are generated
on the bottom side of the bubble and negative (clockwise) vortices are
generated on the top side of the bubble. 
An opposite scenario is observed for the heavy bubbles, where vortices
with positive sign are generated on the top and negative sign on 
the bottom of the bubbles. In later times 
(see especially the SF$_6$ evolution in Fig.~\ref{Fig17}), 
a dilation of the vorticity torus induced by its spiral effect can 
be observed. Looking at the values of the calculated circulation one can 
conclude that for the Atwood numbers of relatively high absolute values 
the vorticity generation rate becomes higher. When the Mach number 
is increased the value of the total circulation is also higher 
as its growth rate during the shock-bubble interaction process 
becomes faster. The effect of the vorticity on the N$_2$ bubble 
is negligible owing to the small differences in densities and acoustic 
impedance between N$_2$ and the surrounding air.

\subsection{Influence of the heat capacity ratio $(\gamma)$ 
on the bubble compression}

In addition to the essential effect of the density ratio across 
the interfaces and the corresponding acoustic impedance difference, 
there is another fundamental parameter that contributes to the interface
evolution. The heat capacity ratio $\gamma$ influences bubble 
compression and deformation. 
This parameter monitors how compressible the medium is. 
For example, SF$_6$ with $\gamma = 1.08$ is much more compressible
than the other mono and di-atomic gases, such as helium and nitrogen 
discussed in the previous section. In most of the literature concerned 
with the shock-bubble interaction problem, the analysis and discussions 
of this phenomenon is focused on the role of the acoustic impedance 
and the pressure misalignment on the interface deformation and vorticity 
production. The effect of $\gamma$ was not highlighted. 

A new hypothetical shock-bubble interaction case study is designed 
to address the role of $\gamma$. This case study considers 
a planar shock of ${\it Ma} = 1.5$ propagating in ambient air 
and interacting with an air bubble characterised by the same densities, 
i.e. zero Atwood number, but different values of $\gamma$. That is
$\gamma=1.2,~1.4$ and $1.67$. 
The physical domain, Fig.~\ref{Fig5}, and computational set-up are 
the same as in subsection~\ref{Layes_validation}.

\begin{figure}[!ht]
\centering
\includegraphics[width=0.235\textwidth]{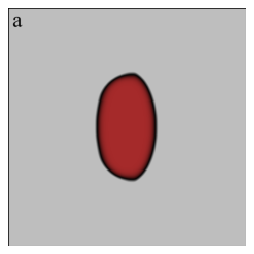}
\includegraphics[width=0.235\textwidth]{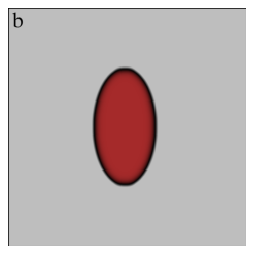}
\includegraphics[width=0.235\textwidth]{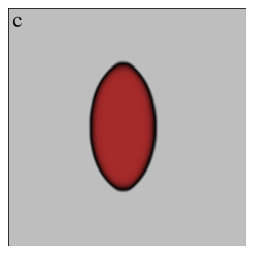}
\includegraphics[width=0.235\textwidth]{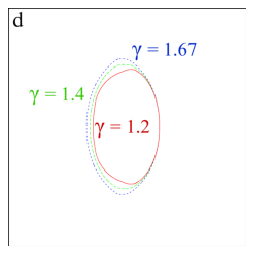}
\caption{Volume fraction contours of the air bubbles 
at time $t = 306~\mu{\rm s}$ for different interaction scenarios: 
(a) $\gamma = 1.2$ (b) $\gamma = 1.4$ (c) $\gamma = 1.67$ 
and (d) comparison of volume fraction profiles.}
\label{Fig21}
\end{figure}

Figures~\ref{Fig21}(a)~to~(c) show the volume fraction contours 
at the same physical time $306~\mu s$ from the beginning of 
the interaction for these three values of $\gamma$. The bubbles 
underwent a compression process as in the case of air/N$_2$ that 
was shown previously. However, the contour of each bubble has 
a slightly different shape. Figure~\ref{Fig21}(d) summarises 
the relative change of these different shapes of the bubbles 
with respect to $\gamma$. While for the case of 
$\gamma = 1.2$ the bubble compresses more than the other 
two bubbles, the bubble with the largest gamma stretches 
vertically more than the others as shown in Fig.~\ref{Fig20}(d).

\begin{figure}[!ht]
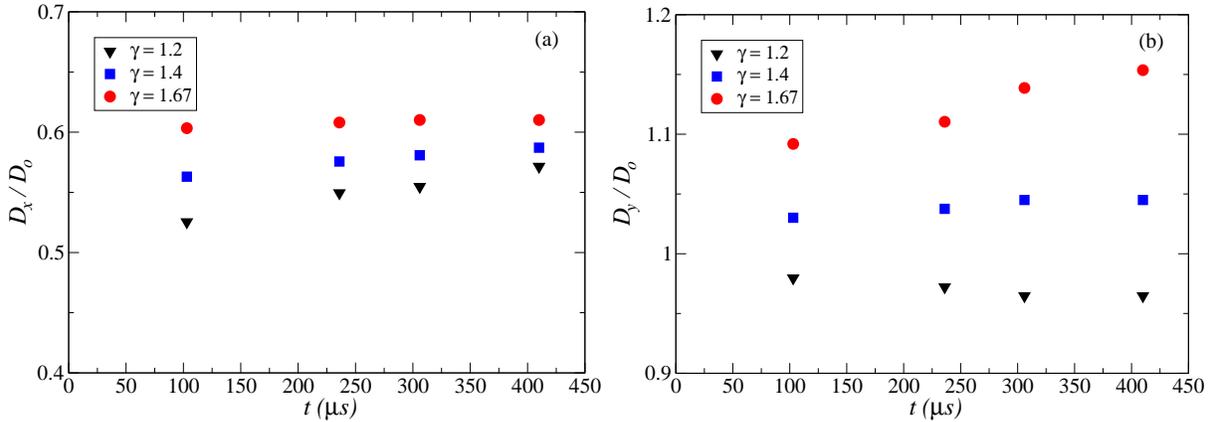

\centering
\includegraphics[width=0.48\textwidth]{Fig22a.eps}
\includegraphics[width=0.48\textwidth]{Fig22b.eps}\\
\caption{Relative change of the bubble size as a function of time 
(a) horizontal diameter 
(b) vertical diameter, 
during the interaction scenarios 
for different heat capacity ratios, $\gamma$.}
\label{Fig22}
\end{figure}

Figure~\ref{Fig22} quantifies the variation of the size of 
the bubbles along their horizontal and vertical diameters, 
with respect to time. $D_0$ is the initial diameter and 
$D_x$ and $D_y$ represent bubble horizontal and vertical 
diameters during the process of the shock-bubble interaction. 
The numerical results confirm 
that the heat capacity ratio $\gamma$ has an effect 
on the interface deformation. However, one has to remember that
using this parameter separately in the discussion can be
misleading. This is because the heat capacity ratio effects 
are already indirectly included in the acoustic impedance since 
the calculation of the sound speed of the gases across 
the interfaces requires $\gamma$. 

The total circulation values recorded for various $\gamma$
are listed in Table~\ref{Table10}. Although these values are 
negligible owing to small difference in acoustic impedance, 
they confirm the observations discussed 
in section~\ref{Mach_Atwood_comparison}.  

\begin{table}
\caption{\label{Table10}Circulation {$\Gamma$(m$^2$/s)} 
as a function of time for various $\gamma$.}
\begin{ruledtabular}
\begin{tabular}{rccc}
Time ($\mu s$) & $\gamma = 1.2$ & $\gamma = 1.4$ & $\gamma = 1.67$\\
\hline
20   & $-0.068$ & 0.000 & 0.080\\
47   & $-0.163$ & 0.000 & 0.185\\
60   & $-0.206$ & 0.000 & 0.233\\
103  & $-0.268$ & 0.000 & 0.271
\end{tabular}
\end{ruledtabular}
\end{table}

\section{\label{sec:5}Conclusions}   
The computations of flows in inhomogeneous media of various physical
regimes leading to shock-bubble interactions were performed using 
a newly developed numerical code based on an Eulerian multi-component
flow model. The numerical approach was validated using available data
from shock tube experiments, for which very 	good qualitative and
quantitative agreements were found. The present numerical approach could
be applied to design better shock-induced mixing processes. In order to 
better understand the bubble shape changes and describe the mechanism
of its interface deformation, the study was extended to include additional
cases for which experimental data cannot be collected. These enabled us
to account for the effect of the Atwood number and shock wave intensity
(various Mach numbers) on the interface evolution and on the vorticity
generation within the surrounding medium. The constant Mach number 
comparison showed that the Atwood number increase leads to higher 
vorticity generation
and its effect on the interface evolution becomes more pronounced. 
Similarly the constant Atwood number comparison shows that increasing
the Mach number produces a higher circulation which also means a higher 
vorticity generation. Apart from highlighting the cases characterised 
by the difference in acoustic impedance the study was extended to account
for the influence of the heat capacity ratio $\gamma$ of the heterogeneous 
media on the interface deformation. The results of this study, which could 
potentially constitute a benchmark test for other numerical simulations, 
confirm that the baroclinic term in the vorticity transport equation 
has a large effect on the interface evolution and the vorticity generation. 
The 2D simulations can only be used as a platform for the analysis of early 
stages of a shock wave-spherical bubble interaction. For longer time 
periods after a planar shock-inhomogeneity interaction, the flow becomes 3D 
and vorticity structures are influenced by vortex stretching.

\bibliography{PRE_jor.bib}
\end{document}